\documentclass[12pt]{article}

\usepackage{arxiv}

\usepackage[utf8]{inputenc} % allow utf-8 input
\usepackage[T1]{fontenc}    % use 8-bit T1 fonts
\usepackage{hyperref}       % hyperlinks
\usepackage{url}            % simple URL typesetting
\usepackage{booktabs}       % professional-quality tables
\usepackage{amsfonts}       % blackboard math symbols
\usepackage{nicefrac}       % compact symbols for 1/2, etc.
\usepackage{microtype}      % microtypography
\usepackage{graphicx}
\usepackage{natbib}
\usepackage{doi}
\usepackage{amsmath}
\usepackage{xcolor}

\hypersetup{
    colorlinks,
    linkcolor={red!50!black},
    citecolor={blue!50!black},
    urlcolor={blue!80!black}
}

\title{Neutral Diversity in 
Experimental Metapopulations 
}

\author{ \href{https://orcid.org/0000-0003-3720-9089}{\includegraphics[scale=0.06]{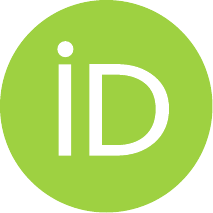}\hspace{1mm}Guilhem Doulcier}\\%\thanks{Use footnote for providing further
	Macquarie University, Sydney, Australia\\
	MPI for Evolutionary Biology, Plön, Germany\\
	\texttt{guilhem.doulcier@normalesup.org} \\
	\And
	\href{https://orcid.org/0000-0002-7248-9955}
 {\includegraphics[scale=0.06]{orcid.pdf}\hspace{1mm}Amaury Lambert} \\
	SMILE -- Stochastic Models for the Inference of Life Evolution\\ Institut de Biologie de l'ENS (IBENS),\\ \'Ecole Normale Supérieure,\\ CNRS UMR8197, INSERM U1024\\
 \&\\
 Centre Interdisciplinaire de Recherche en Biologie (CIRB),\\ Coll\`ege de France, CNRS UMR7241, INSERM U1050\\
 PSL Universit\'e, Paris, France\\
 \texttt{amaury.lambert@ens.fr} \\
}

\date{}

\renewcommand{\headeright}{Preprint}
\renewcommand{\undertitle}{Preprint}
\renewcommand{\shorttitle}{Neutral Diversity in %Experimentally Nested Populations 
Experimental
Metapopulations}

\newcommand\iid{i.i.d.~}
\newtheorem{prop}{Proposition}
\newtheorem{proof}{Proof}

\hypersetup{
pdftitle={\shorttitle},
pdfauthor={Guilhem Doulcier, Amaury Lambert},
}
\renewenvironment{proof}[1][\proofname]{{\large{\bfseries Proof of #1}}\par}{\hfill$\square$}

\begin{document}
\maketitle

\setlength{\headheight}{14.5pt}

\begin{abstract}
	New automated and high-throughput methods allow the manipulation and selection of numerous bacterial populations. 
	In this manuscript we are interested in the neutral diversity patterns that emerge from such a setup in which many bacterial populations are grown in parallel serial transfers, in some cases with population-wide extinction and splitting events. We model bacterial growth by a birth-death process and use the theory of coalescent point processes.
	We show that there is a dilution factor that optimises the expected amount of neutral diversity for a given number of cycles, and study the power law behaviour of the mutation frequency spectrum for different experimental regimes. 
	We also explore how neutral variation diverges between two recently split populations  by establishing a new formula for the expected number of shared and private mutations.  
	Finally, we show the interest of such a setup to select a phenotype of interest that requires multiple mutations.  
\end{abstract}

% keywords can be removed

\section{Introduction}

Microbial communities are ubiquitous in nature and perform critical roles in biogeochemical cycles \citep{fuhrman_microbial_2009}, agricultural productivity \citep{trivedi_response_2016} and human health \citep{pflughoeft_human_2012}.
Microbial diversity has been recognised to have a strong impact on these dynamics \citep{kassen_diversity_2000,maron_high_2018,delgado-baquerizo_lack_2016}.
As a consequence, considerable effort is being deployed in understanding and engineering their eco-evolutionary dynamics by either building ``bottom-up'' synthetic communities \citep{li_design_2022}, screening variants from environmental samples \citep{milshteyn_mining_2014}, or finally developing ``selective breeding'' for microbial communities \citep{mueller_engineering_2015,arias-sanchez_artificially_2019}.
In the following, we use stochastic processes to study some basic parameters of a device that could be used for the screening and selection of microbial communities.
This device is abstract but inspired by recent advances in high-throughput microbial population manipulation methods.

% What is experimental evolution and microfludics 
Microbial populations offer numerous advantages in the domain of experimental evolution (i.e., the study of evolutionary dynamics happening in real time as a response to conditions imposed by the experimenter; \citealt{kawecki_experimental_2012}).
These advantages include large population sizes, easily manipulable environments, the possibility to freeze and store whole populations indefinitely\ldots~Experimental evolution requires the set-up of many parallel bacterial cultures that can take several forms from bottles (in the order of 1 litre, $\approx 1L$) to tubes ($\approx10^{-3}L$), microplates ($\approx  10^{-4}L$), or microfluidic compartments ($\approx 10^{-9}L$). 
Recently, new techniques for the high-throughput manipulation of bacterial populations have emerged. 
For instance, digital millifluidics \citep{cottinet_diversite_2013,dupin_cultures_2018, doulcier_dropsignal_2019,ardre_leader_2022} allows producing and imaging thousands of droplets of culture broth within a carrying fluid. 
The droplets amount to around $2\times 10^{-6}L$ with a carrying capacity of $10^5$ to $10^6$ cells \citep{cottinet_diversite_2013}. 
Droplets can be imaged and quantitative measures be performed (e.g., optical density, fluorescence signal) during growth of the bacteria, allowing a high-throughput monitoring of ecological dynamics.

% WHat is a nested population design and why would we want it ? 
Being able to phenotype, screen and eventually artificially select microbial communities at scale could open new possibilities in experimental evolution as well as biological engineering. 
In particular, a new kind of experimental design becomes accessible, compared to more traditional experimental evolution set-ups (e.g.,~screening by plating or flow cytometry). 
We call this design ``nested populations'', in which both cells and microbial populations (e.g., physically bounded by droplets) have their own demography with birth and death events.
The consequence of such a nested design is that populations themselves become units of selection \citep{lewontin_units_1970} in their own right. 
Such experiments are routinely performed in microcosms \citep{hammerschmidt_life_2014}. 
However, the ability of millifluidic devices to monitor in the order of a thousand cultures and retrieve some of them for analysis makes them particularly suitable for the artificial selection of microbial communities \citep{xie_steering_2021}, such as through ecological scaffolding \citep{doulcier_eco-evolutionary_2020, black_ecological_2020}. 

% Question - Outline of the manuscript.
Neutral diversity in experimentally nested populations is the focus of this manuscript.
The aim is to build a quantitative understanding of simple diversity patterns within the experimental set-up. 
The main objective is to understand how they can be manipulated by changing the parameters that are accessible to the experimentalist or device designer. 
We consider the hypothesis that a high neutral diversity is favourable for the experiment, either because it increases the probability to screen individual variants of interest, or because populations with high diversity are more likely to have properties of interest themselves. 
Moreover, neutral diversity offers a solid foundation for more complex models of microbial communities \citep{ofiteru_combined_2010}.
First, a model of the device is presented. 
It relies on the assumption that cells undergo constant exponential growth. 
The optimal operating regime parameters of the machine (e.g.,~dilution factor, duration of droplet growth cycles, carrying capacity) are derived from characteristics of the biological material: birth and death rates. 
From a theoretical perspective, the system constitutes a dynamical meta-population in discrete space with explicit demography. 
It contrasts with simpler models, in which demography is simplified \citep{etheridge_drift_2008}, as well as with more complex spatially structured models \citep{barton_neutral_2002,barton_modelling_2013}, in which space is continuous.
Second, a coalescent model of the population across bottlenecks is proposed and coupled with a neutral mutation model with infinite alleles. 
This allows computation of the number of mutations and the distribution of allele frequencies within droplets after several droplet growth cycles.
It shows that small bottlenecks are required to maximise diversity in one cycle, but larger bottlenecks are more favourable for diversification across many cycles. 
Then, the effect of splitting a droplet into several lineages is studied by computing the number of mutations accumulated in a single lineage or all the droplet lineages.
Finally, a simple mutation accumulation model illustrates the interest of droplet-level selection for artificial selection.

\section{Modelling Nested Population Dynamics}

\begin{figure}
\begin{center}
\includegraphics[width=\linewidth]{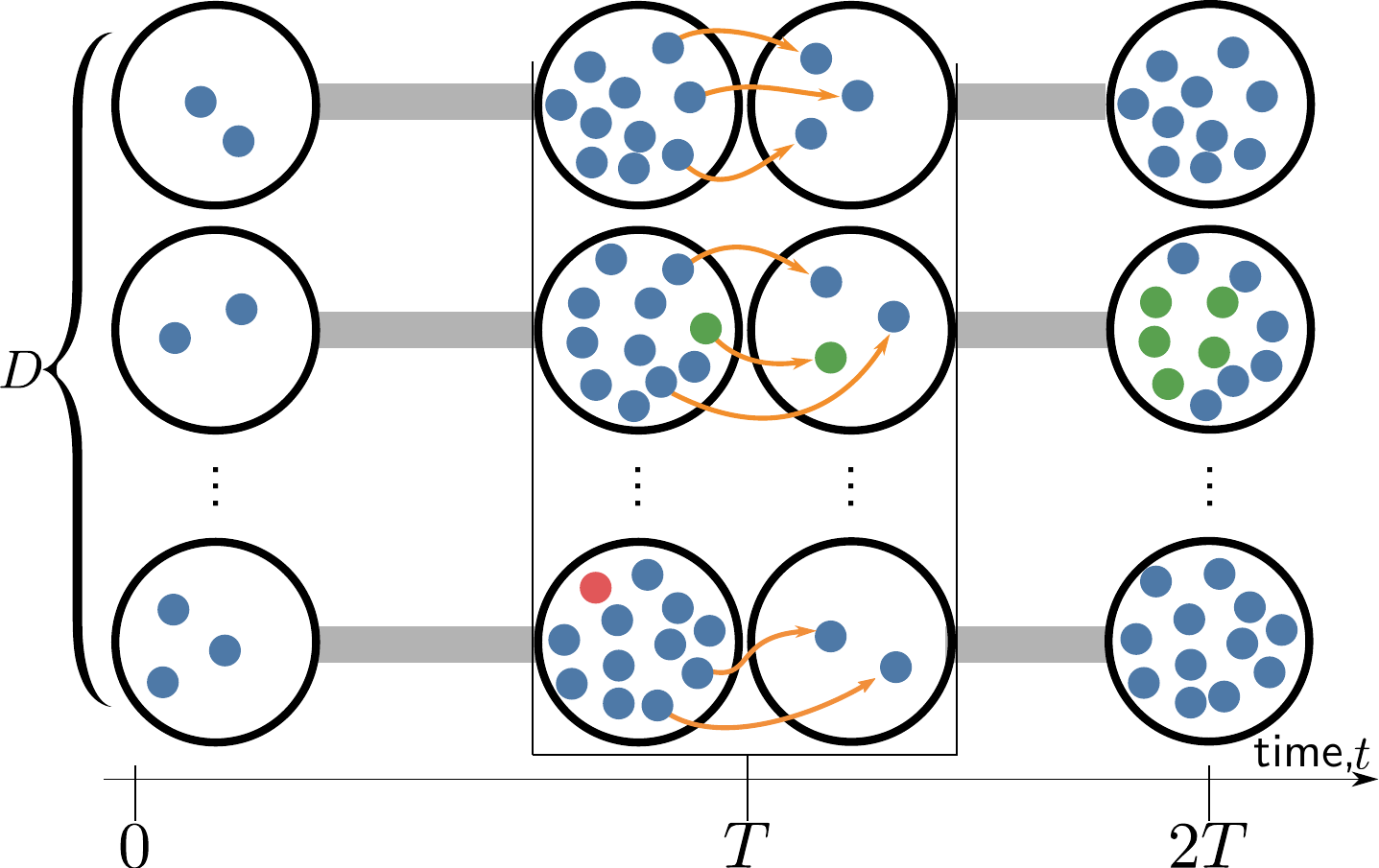}
\caption{\label{fig:machine}
\textbf{\textbf{Sketch of the experimental set-up.}} 
Let a population of cells following a birth-death process with rates $b,d$ be distributed within $D$ droplets. 
After a growth phase duration $T$, a new cycle starts: the content of each droplet is diluted to seed a new droplet. 
Each cell has a probability $\delta$ to be transferred at the start of the next cycle. 
Here, the \emph{serial transfer} regime is depicted: each droplet is diluted into exactly a single new droplet in the next cycle. 
In the full \emph{nested populations'} design, a droplet can be split into several droplets in the cycle or removed altogether.}
\end{center}
\end{figure}

Consider a device that allows the manipulation of droplets via serial transfers (Figure \ref{fig:machine}). 
Cells are distributed among a train of $D$ droplets (called populations or droplets). 
The birth and death of cells are modelled by a linear branching process with constant rates $b$ for birth and $d$ for death. 
The net growth rate $r:=b-d$ is called the Malthusian parameter. 
After a duration $T$, a new train of $D$ droplets is prepared by diluting them $\frac 1\delta$ fold.
Hence, for each dilution event, a cell has a probability $\delta$ of being sampled and thus being present in the new droplet. 
This procedure is repeated periodically; each dilution followed by a growth phase constitutes a \emph{cycle} of the experiment or a \emph{droplet generation}.

More formally, the initial population contains $Z_0 = c$ cells and is submitted to a bottleneck with sampling probability $\delta$ at the beginning of every cycle except the first, meaning that bottlenecks occur at times $T,2T \ldots nT$. 
The $n$th cycle corresponds to the slice of time $[(n-1)T,nT)$. Thus, ``the end of cycle $n$'' corresponds to the moment $nT^{-}$, just before the $n$th dilution. To illustrate this, at the end of the third cycle, $t=3T^{-}$, the population has experienced two bottlenecks at time $T$ and $2T$. 

Birth $b$ and death $d$ rates depend on the biological material used (e.g., species, strain), as well as the culture medium, and are not easily controlled. 
However, the duration of the growth phase $T$, the dilution factor $\delta$, and the number of droplets $D$ can be changed by altering the experimental set-up. 
A model can help to predict the effect of those parameters and determine those that should be the focus of engineering efforts, such as the development of devices supporting larger droplets, an increased number of same-size droplets, a wider range of dilution factors, or droplets that are stable for longer times.

%\begin{table}[!h]
%	\caption{Parameter names and symbols reference}
% 	\centering
%\begin{tabular}{ll|ll}
%& \textbf{droplet-level parameters} & & \textbf{cell-level parameters}\\
%\hline
%$D$ & Population size & $b$ & Birth rate\\
%$T$ & Cycle duration & $d$ & Death rate\\
%$n$ & Number of cycles & $r$ & Malthusian parameter (b-d)\\
%$K$ & Carrying capacity & $\delta$ & Survival probability  (at a bottleneck) \\
%$c$ & Initial number of cells & $\theta$ & Mutation rate \\
%\end{tabular}
%\end{table}

\subsection{Optimal Operating Regime}
\label{sec:optimal}

When designing a serial transfer experiment, the operator has three main parameters that might be controlled: 
the size of the cultures (and, by extension, the carrying capacity of droplets, $K$, in number of cells), 
the duration of the growth phase separating two successive transfers $T$, 
and the dilution factor $\delta$. 
Two problems must be avoided: if population sizes are too small and dilution too high, the resulting cultures might be empty.
Conversely, if the population sizes are too large, and dilution too low, the population will spend most of its time in a stationary phase, with little effect of bottlenecks.

Any dilution event presents the risk of extinguishing the population. 
When performing a serial transfer experiment, this must be avoided at all costs because an empty microcosm brings about the end of the experiment (at least for the given independent lineage).
In a nested populations' design, the presence of some empty microcosms can be tolerated 
because empty patches in the population can be filled by splitting a single parent droplet into several offspring droplets in the next generation.

In general, the stationary phase is not desirable for several reasons. 
First, a population that reaches saturation will go through fewer generations than if it was growing freely, reducing the potential evolutionary dynamics. 
Moreover, physiological changes in the stationary phase might result in undesired phenotypic effects on the population. 
Finally, in the case of millifluidic experiments, saturating densities are known to increase the risk of cross-contamination between droplets.

For all these reasons, there is an optimal dilution factor that keeps the population in an exponential phase while maximising the population size, 
at which selection experiments should be conducted.

\begin{figure}
\begin{center}
\includegraphics[width=0.5\textwidth]{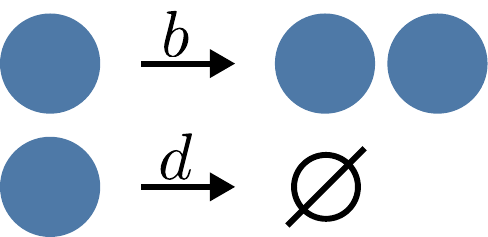}
\caption{\label{fig:linear_bd_events}
\textbf{Events in a Linear-Birth-Death Model.} Individuals give birth to new individuals at a constant rate $b$, and die at constant rate $d$, independently. The process is super-critical if $b>d$.}
\end{center}
\end{figure}

A model of population dynamics can provide a first estimate of the optimal range of parameters for an experiment.
In the following, a stochastic model of cells in exponential growth conditions (i.e.,~super-critical) with periodic bottlenecks is used to derive the probability of losing a single cell lineage, or a single droplet lineage due to the effect of dilution, as a function of experimentally accessible parameters.
Saturation phenomena are not modelled explicitly because the birth and death rates are considered independent of population size,
but the population dynamics are required to stay under a carrying capacity threshold.

\begin{figure}
\begin{center}
\includegraphics[width=0.5\textwidth]{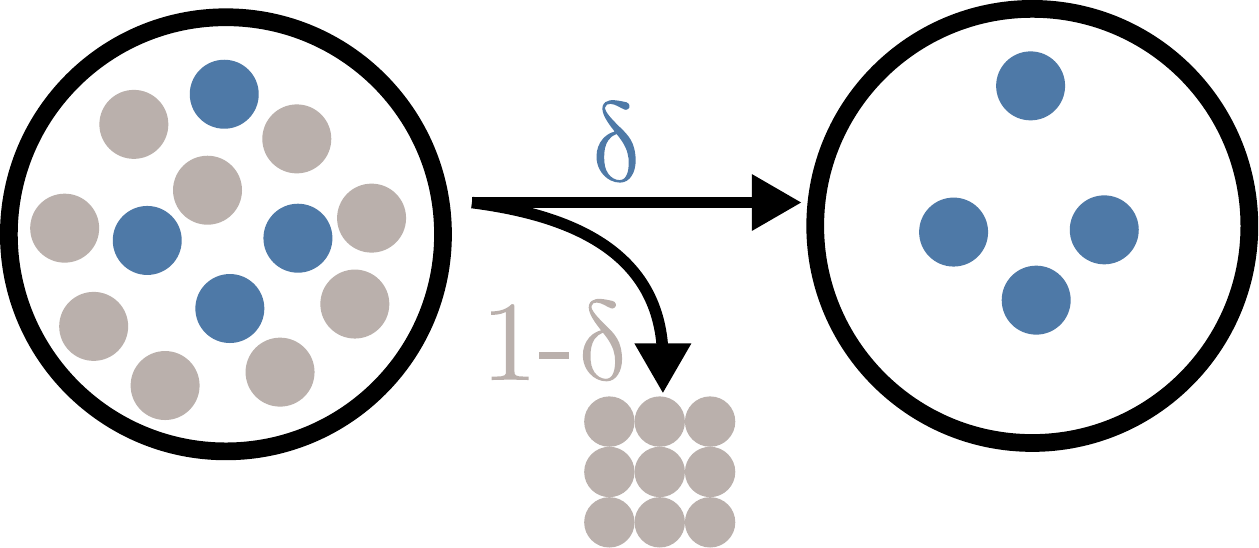}
\caption{\label{fig:dilution_process}
\textbf{Dilution process.} Individual cells are independently selected to be transferred to the next cycle (with probability $\delta$) or to be discarded (with probability $1-\delta$).}
\end{center}
\end{figure}

\subsubsection{Survival of a single lineage}

A first quantity that can be derived from the linear branching process with periodic dilution that models the population dynamics is the probability that a single initial cell has no descent in the population after $n$ cycles.

\begin{prop}[Survival Probability]
\label{lem:survival_prob}

Cells within droplets in serial transfers are modelled by a linear birth-death process, with constant parameters $b$ and $d$, that is subject to periodic bottlenecks $\delta$ every duration $T$.

The probability $s_n$ that a lineage spawned by a single cell at the beginning of the first cycle is not extinct at the end of the $n$th cycle is:
\begin{equation}
s_n = 1 - h(Q_1Q_\delta^{n-1},0),
\end{equation}
where $s \mapsto h(M,s)$ is the linear fractional function with coefficient $M \in M_2(\mathbb R)$:
\begin{equation}
\label{eq:linear_fractional}
\forall M = \begin{bmatrix} v & w \\ x & y \end{bmatrix} \in M_2(\mathbb R),\quad h(M,s) = \frac{vs+w}{xs +y}.
\end{equation}
and for $\varepsilon \in (0,1]$ the matrix $Q_\varepsilon$ is:
\begin{equation}\label{eq:Qdelta}
Q_\varepsilon = \begin{bmatrix}
p-1+\varepsilon (1-q)& (1-\varepsilon)+\varepsilon q \\
p-1 & 1
 \end{bmatrix}
\end{equation}
%\begin{equation*}
%Q_0 = \begin{bmatrix}
%\delta(p-q) & q \\
%\delta(p-1) & 1
% \end{bmatrix};
%\;\;\;
%Q = \begin{bmatrix}
%p-(1-\delta)-\delta q& (1-\delta)+\delta q \\
%p-1 & 1
% \end{bmatrix}
%\end{equation*}
with,
\begin{center}
\begin{tabular}{llll}
\hline
 &  & $p(b,d,T)$ & $q(b,d,T)$\\
\hline
Sub-critical cells & $b < d$ & $\frac{-r}{d-be^{rT}}$ & $\frac{d(1-e^{rT})}{d-be^{rT}}$\\
Critical cells & $b = d$ & $\frac{1}{1+bT}$ & $\frac{bT}{1+bT}$\\
Super-critical cells & $b > d$ & $\frac{re^{-rT}}{b-de^{-rT}}$ & $\frac{d(1-e^{-rT})}{b-de^{-rT}}$\\
\hline
\end{tabular}
\end{center}

\begin{flushright}
(Proof page \pageref{proof:survival_prob}.)
\end{flushright}
\end{prop}

\begin{figure}
\begin{center}
\includegraphics[width=0.5\textwidth]{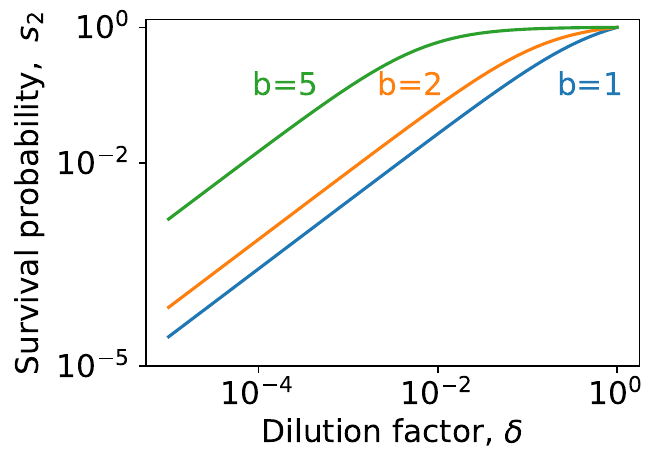}
\caption{\label{fig:survival_s1_f_delta}
\textbf{Survival probability at the end of the second cycle $s_2$.} This corresponds to a first growth phase, a dilution event, and a second growth phase.  
The probability is presented as a function of the dilution factor $\delta$ for pure birth processes $b>0, d=0, T=1$.}
\end{center}
\end{figure}

Proposition \ref{lem:survival_prob} shows that the survival probability of a lineage depends on the birth rate $b$ and death rate $d$ of the cells but is also a function of the dilution factor $\delta$, duration of the growth phase $T$, and the number of cycles $n$. When considering a single dilution and a pure birth process ($s_2$, Figure \ref{fig:survival_s1_f_delta}), the survival probability is equivalent to $\delta \mathbb E(Z_T)$ when $\delta$ is small; hence the linear increase with slope $1$ in log-log scale.

The numerical computation of $Q_\delta^{n-1}$ might be problematic due to repeated multiplication of small numbers.
However, since the final result only involves the ratio $\left(Q_\delta^{n-1}\right)_{01}/\left(Q_\delta^{n-1}\right)_{11}$, it is possible to normalise $Q_\delta$ to have its smallest entry as $1$. Indeed, this ratio does not depend on a multiplicative scalar on the matrix: $\forall \alpha>0$, $h(Q_\delta^n,0) = h(\alpha Q_\delta^n,0)$. Taking $\alpha=\frac{1}{\delta}$ greatly improves the numerical stability of the computation.

The limit of this probability when the number of cycles increases gives a clearer understanding of the long-term behaviour of the population:

\begin{prop}[Long-Term Survival Probability]
\label{lem:survival_prob_long_term}
Let $s_n$ be the survival probability after $n$ cycles of the lineage spawned by a single cell.

\begin{equation*}
\lim\limits_{n\to +\infty} s_n =
\begin{cases}
 0 &\text{if }  \delta e^{rT}\le 1\\
 \frac{r(\delta-e^{-rT})}{\delta b (1-e^{-rT})} &\text{otherwise.}
\end{cases}
\end{equation*}

\begin{flushright}
(Proof page \pageref{proof:survival_prob_long_term}.)
\end{flushright}
\end{prop}

\begin{figure}
\begin{center}
\includegraphics[width=0.5\textwidth]{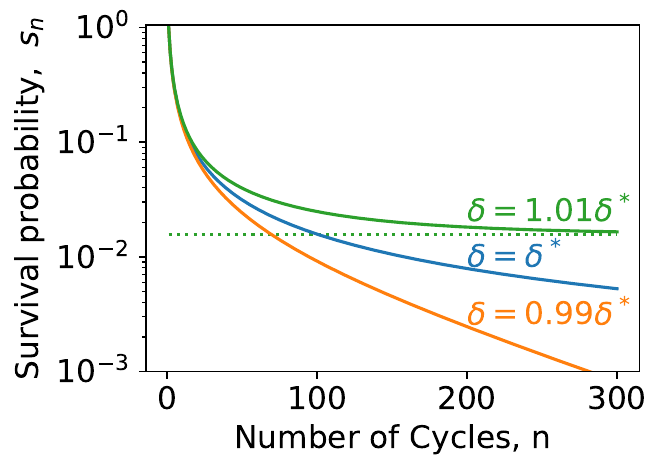}
\caption{\label{fig:survival_sn_f_n}
\textbf{Survival probability for several cycles $s_n$.}
The probability is presented for pure birth processes $b=1, d=0, T=1$. Above the critical threshold $\delta^*$, the survival probability does not tend towards $0$. 
The dotted line corresponds to the limit $s_\infty$.}
\end{center}
\end{figure}

Proposition \ref{lem:survival_prob_long_term} confirms that, in the long run, lineages become extinct with certainty ($ s_\infty=0$) if and only if the expected number $\delta e^{rT}$ of cells descending from a single initial cell and surviving the first bottleneck is smaller than $1$. It gives the survival probability otherwise (see Figure \ref{fig:survival_sn_f_n}).

In the following, only super-critical populations will be considered $(b>d)$.

\subsubsection{Optimal cycle duration and dilution}

Saturation of cell dynamics is not desirable, as mentioned earlier.
Depending on the nature of the cells (e.g., species, strain), and of the medium (e.g., pH, nutrient availability, temperature), it is possible to define an experimental carrying capacity $K$ that corresponds to the number of cells that can be sustained in a droplet without saturation.
The simple linear birth-death model cannot represent saturating populations because no density dependence is included in this model; thus, for the model to be coherent, the duration of the growth phase must be short enough that the population size does not reach the carrying capacity:

\begin{prop}[Maximal Cycle Duration]
\label{lem:tstar} Let $T^*$ be the maximal cycle duration before
reaching saturation. Cells are following a super-critical birth-death process
with growth rate $b-d=r>0$. The carrying capacity is $K$ and the initial
number of cells is $c$ :
\begin{equation}
T^* = -\frac{\ln(\frac cK)}{r}
\end{equation}

\end{prop}

\begin{figure}\begin{center}
\includegraphics[width=0.5\textwidth]{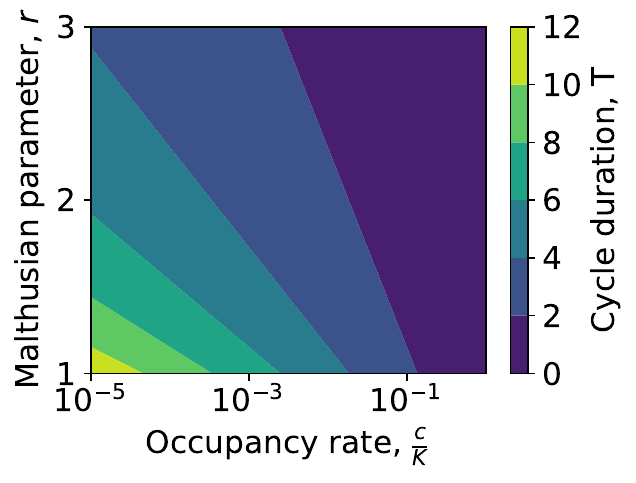}
\caption{\label{fig:criticalT}
\textbf{Maximal Cycle Duration $T^*$} as a function of the Malthusian parameter $b-d$ and the initial occupancy $\frac{c}{K}$.
}\end{center}\end{figure}

Proposition \ref{lem:tstar} shows that the optimal duration of the growth phase is linear in the inverse of the Malthusian parameter $r$ of the population (Figure \ref{fig:criticalT}), meaning that a population that grows (on average) twice as fast as another should be subject to cycles half as long as the other, for a given initial occupancy $\frac cK$.

Additionally, the optimal duration of the growth phase is proportional to the logarithm of the initial occupancy $\frac cK$ of the droplet (with a minus sign, since this logarithm is always negative or zero as $c\leq K$). As a consequence, for a given strain, multiplying the volume of the droplets by two, or dividing the inoculum size by two will increase the maximal duration of the growth phase by $\frac{\ln{2}}{r}$.

To keep the same maximal duration $T^*$ if the  Malthusian parameter $r$ is doubled, the carrying capacity of the droplet must be multiplied by the inverse of the previous initial occupancy $K/c$.

This result holds for a single cycle only. For a given dilution factor $\delta$, the population is shrunk by an expected ratio $\delta$, while for a given cycle duration $T$, the population is expanded by an expected ratio $\delta e^{rT}$. To prevent the population from saturating for all cycles, the initial occupancy at the beginning of each cycle must be constant. This consideration allows discovery of the optimal dilution factor $\delta^*$ when the growth phase duration is fixed. 
By simply setting $\delta^*$ to be the solution of $\delta e^{rT}=1$, we obtain $\delta^*=e^{-rT}$. We record this result in proposition \ref{lem:dstar}.

\begin{prop}[Optimal dilution factor]
\label{lem:dstar} Let $\delta^*$ be the optimal dilution factor for
which the expected number of cells is constant across
generations. For cells following a birth-death process
with growth rate $b-d=r>0$ and a growth phase duration $T$:
\begin{equation}
\delta^* = e^{-rT}.
\end{equation}

If $T=T^*$ (Proposition \ref{lem:tstar}),
\begin{equation}
\delta^* = \frac cK
\end{equation}
\end{prop}

Proposition \ref{lem:dstar} shows that the dilution sampling probability should be equal to the initial occupancy when the duration of the cycle is maximal.

To summarise, the optimal operating regime of the experiment can be expressed from the Malthusian parameter of the population and the initial occupancy of the droplets $\frac cK$.
As a result, the dilution sampling probability is $\delta^* = \frac{c}{K}$, and the duration of a cycle is $T^* = -\frac{\log(\delta^*)}{r}$.
Fixing any two of $(c,K,T,\delta)$ values constrains the other two.

When the experiment is in the optimal regime, the expression of the survival of a lineage at cycle $n$ is simpler:

\begin{prop}[Optimal Regime Survival Rate]
\label{lem:snstar}
Let $s_n^*$ be the survival probability of a lineage after $n$ cycles of duration $T$ and with bottleneck $\delta^*=e^{-rT}$, where $r=b-d>0$ is the Malthusian parameter of the population.

Then:
\begin{equation}
\label{eqn:survival-in-optimal-regime}
s_n^* = \frac {r}{bn(1-\delta^*)+\delta^* r}
\end{equation}

\begin{flushright}
(Proof page \pageref{proof:snstar}.)
\end{flushright}
\end{prop}

In the optimal regime, each initial cell has, on average, one descendant cell surviving the next bottleneck. 
The process counting the number of cells at time $kT$ is thus a critical Galton--Watson process (as a function of $k$).
Proposition \ref{lem:snstar} shows that, in the optimal regime, the survival probability of a lineage decreases as the inverse of the number of cycles, which is typical of critically branching populations.

Additionally, when taking a finite number of cycles $n$, the survival does not tend towards zero, even for vanishingly small bottlenecks $\delta$.
This derives from the fact that, in the optimal regime, a small bottleneck is compensated by a long cycle duration $T^*$; thus, vanishingly small bottlenecks correspond to infinitely long cycles.

Finally, in the case of pure birth (i.e., $d=0$), the survival of a lineage is independent of the birth rate $b$.
It is certain if there is no bottleneck ($\delta=1$), and tends toward $\frac{1}{n}$ for vanishingly small bottlenecks ($\delta \to 0$).\\

%% Transition
Overall, once the size of the droplets (which constrains $K$) and the initial occupancy (which constrains $c$) have been chosen by the operator, 
other parameters of the machine (duration of the growth phase $T$, dilution factor $\delta$) can be deduced---and, conversely, fixing $T$ and $\delta$ constrains $K$ and $c$.  
The next section explores how one should select these parameters when the aim is to maximise genetic diversity within and between the droplets.

\section{Modelling Neutral Diversity}

Neutral diversity concerns mutations arising in the population of cells that are assumed not to change their birth or death rates. 
Neutral diversity gives rise to recognisable \emph{patterns} that can be predicted from a mechanistic model of birth-death in the population. 
In experimental evolution, and \emph{a fortiori} in artificial selection, it is desirable to increase the diversity within the population because this allows greater exploration of the phenotypic space. Indeed, mutations that are essentially neutral for cells might be of interest for the experimenter, or be intermediate states towards new phenotypes.

In the following, mutations follow a Poisson Point Process with constant rate $\theta$ over the lifespan of the cells, independently of their genealogy. 
As a consequence, the time between two mutations along a lineage (regardless of births and deaths) is exponentially distributed, and thus has no memory: 
the conditional expected time to the next mutation will be the same for all cells, irrespective of their age or the time of the last mutation in the lineage.
This is a simplifying assumption that represents the spontaneous nature of mutations while ignoring the existence of mutations that can change the mutation rate \citep{sniegowski_evolution_1997,sniegowski_evolution_2000}.

\subsection{Coalescence times and the Coalescent Point Process}

\begin{figure}
\begin{center}
\includegraphics[width=\textwidth]{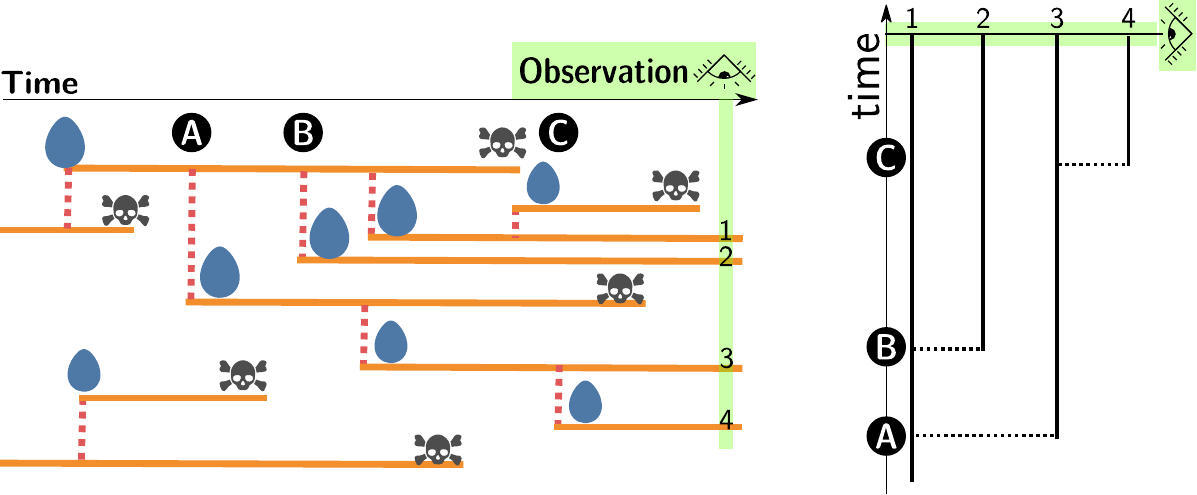}
\caption{\label{fig:bd_to_cpp}
\textbf{\textbf{From a Birth-Death Process to a Coalescent Point Process.}}
On the left-hand side is a birth-death process where a number of individuals give birth (eggs) and die (skulls) at different points in time, which flows from left to right. 
On the right-hand side is the corresponding continuous coalescent tree, where time flows from bottom to top. 
Note that at time $C$, lineage $3$ coalesces with lineage $4$, and that at time $B$, lineages $1$ and $2$ coalesce.
Finally, at time $A$, lineage $(1,2)$ coalesces with lineage $(3,4)$.}
\end{center}\end{figure}

The linear branching process (with constant birth rate $b$ and death rate $d$) yields the full genealogy of the population (Figure \ref{fig:bd_to_cpp}, left).
However, the standing diversity at a given time in the population is affected (i) neither by the mutations occurring on lineages that do not have extant individuals (because their mutations have been lost) (ii) nor by mutations that are ancestral to the whole population (because they are shared by all individuals in the population).
Thus, to characterise the standing diversity, it is sufficient to have knowledge of the \emph{coalescent tree} of the population (Figure \ref{fig:bd_to_cpp}, right),
which is the genealogy of the extant individuals up to their most recent common ancestor.

\begin{figure}
\begin{center}
\includegraphics[width=0.5\textwidth]{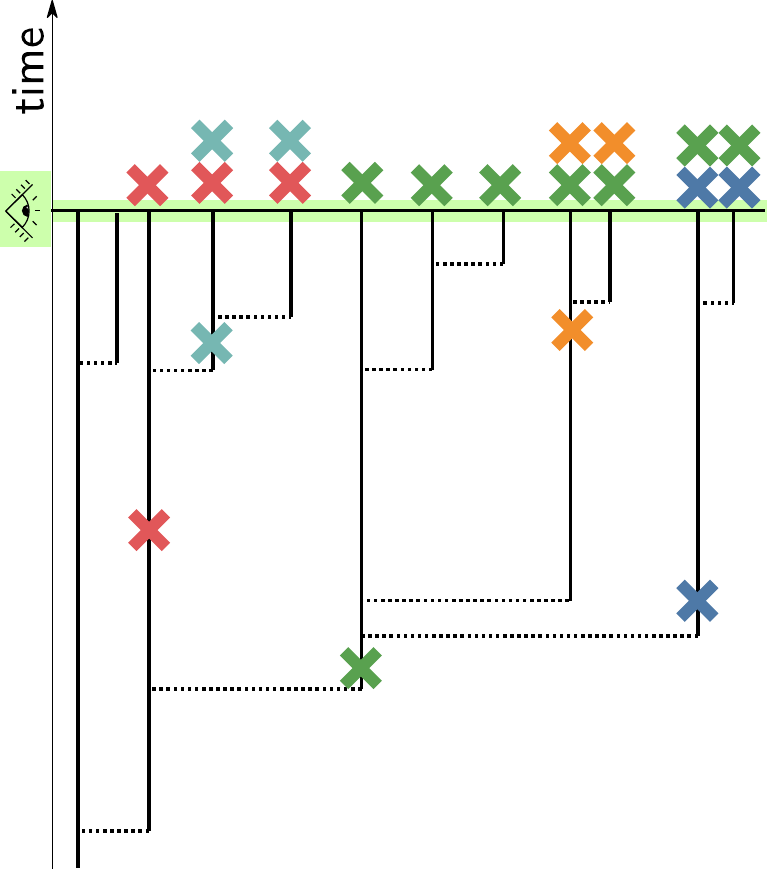}
\caption{\label{fig:cpp_illus}
\textbf{Neutral mutations over the coalescent tree}.
The neutral mutations (coloured crosses) are distributed following a Poisson Process over the (real) coalescent tree (black). Each individual may carry several mutations distinguishing it from the most recent common ancestor.}
\end{center}
\end{figure}

Coalescent Point Processes (CPP) are stochastic processes whose realisations are real trees with the same probability as the coalescent tree of the corresponding branching process \citep{popovic_asymptotic_2004,lambert_birthdeath_2013}. 
A CPP is defined by a time horizon $\tau \in \mathbb R_+$ and a node depth distribution $f_H$.
The CPP is the sequence of independent and identically distributed variables $(H_i)_{i=1\ldots N}$ following $f_H$ and stopped at the first element $N$ such that $H_N>\tau$ (the CPP is said to be ``stopped at $\tau$'').
Usually, the node depth distribution is expressed in the form of the inverse tail distribution $F$:
\begin{equation}
\label{eq:F}
F(t) := \frac{1}{P(H>t)}
\end{equation}

\subsection{Measuring neutral diversity}

Neutral mutations do not affect the genealogy and can thus be superimposed \emph{a posteriori} on the coalescent tree.
Consider that mutations appear following a Poisson point process with constant rate $\theta$ over the coalescent tree. 
Thus, a mutation is a point on the coalescent tree, as illustrated in Figure \ref{fig:cpp_illus}. 
Additionally, assume that reverse mutations are impossible (an assumption referred to as the ``infinite sites model''), 
so that all individuals standing above the mutation in the coalescent tree (i.e., the descent of the mutation point) share the mutation (crosses at the top of Figure \ref{fig:cpp_illus}). 
Individuals may carry zero, one, or several mutations.

The mutational richness of the population $M$ (or total diversity) is the number of unique mutations found in the population.
Its expected value is proportional to the total coalescent tree length.

The mutation frequency spectrum $(a_k)_{k\in\mathbb N}$ is another measure of diversity that counts how many mutations are represented by $k$ individuals in the population.

All these measures require some knowledge of the shape of the coalescent tree of the population.
The next paragraph is dedicated to establishing this for the simple case of serial transfer, while the next section is dedicated to the case of splitting droplets.

\subsection{Diversity Within Droplets in Serial Transfer}

\begin{figure}
\begin{center}
\includegraphics[width=0.5\textwidth]{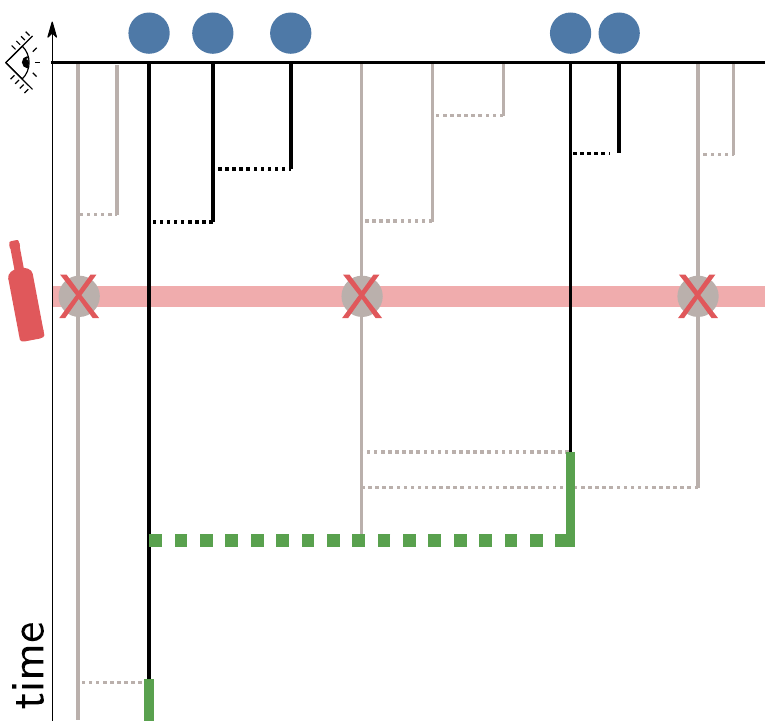}
\caption{\label{fig:thinning}
\textbf{Bottlenecks are modelled by thinning the process}.
A bottleneck at some point in the past (red bottle) results in the extinction of some lineages (in grey), which modifies the genealogy (thick green lines).}
\end{center}
\end{figure}

Establishing the law of the CPP of a lineage within serial transfer requires identification of the law of the branch length. 
This law is well known for simple branching processes such as the linear birth-death process with parameters $(b,d)$ (\cite{lambert_birthdeath_2013}, Proposition 5).

The addition of repeated bottlenecks with period $T$ is also possible within the theory (\citep{lambert_birthdeath_2013}, Proposition 7) by thinning the original process (Figure \ref{fig:thinning}). 
Each bottleneck at time $iT$, $i=1,\ldots,n$ may remove independently each branch of the CPP with probability ($1-\delta$) (in grey in Figure \ref{fig:thinning}).
Removing a branch in the past (at time $iT$) may result in removing several branches in the present, and require an adjustment to branch length (green in Figure \ref{fig:thinning}).
The number of branches removed and the adjustment to the branch length distribution can be computed from the law of the branch length of the CPP in the absence of the bottleneck, the sampling probability $\delta$, and the period of the bottleneck $T$.
As a result:

\begin{figure}
\begin{center}
\includegraphics[width=0.5\textwidth]{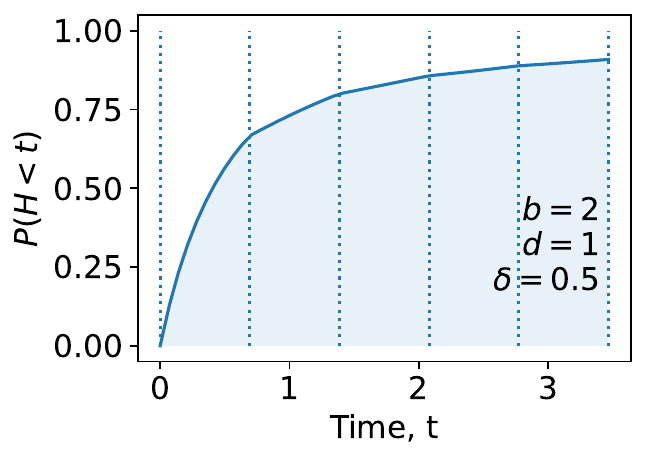}
\caption{\label{fig:cdf_H}
\textbf{Cumulative probability for the branch length}.
Dotted lines represent bottlenecks.}
\end{center}
\end{figure}

\begin{prop}[Coalescent tree of a lineage] \label{lem:coal_lineage}
Let $\mathcal T_n$ be the random coalescent tree spawned by a single cell with extant descent at the end of the $n$th cycle.
Then
$\mathcal T_n$ is a CPP stopped at $nT$, with inverse tail distribution $F$:
\begin{equation*}
     \forall t = kT+s, \quad  F(t) = \delta^k F(t) + (1-\delta)\sum_{j=1}^k\delta^{j-1}F(jT) = 
1 + \frac br \left[ e^{rs}(\delta e^{rT})^k -1
+  (1-\delta) e^{rT} \frac{1-(\delta e^{rT})^k}{1-\delta e^{rT}}   \right ]
\end{equation*}
with $b>0$, $d\geq0$, $b\neq d$, $k\in \mathbb N$ and $0\leq s<T$. Additionally, in the case of critical dilution ($\delta^* = e^{-rT}$), we have:
\begin{equation*}
\forall t = kT+s, \quad
F(t) = 1 + \frac br \left (e^{rs}-1 +  k(e^{rT}-1) \right ).
\end{equation*}

The case $b=d$ is covered in the appendix.

\begin{flushright}
(Proof page \pageref{proof:coal_lineage}.)
\end{flushright}
\end{prop}

\begin{figure}
\begin{center}
\includegraphics[width=0.5\textwidth]{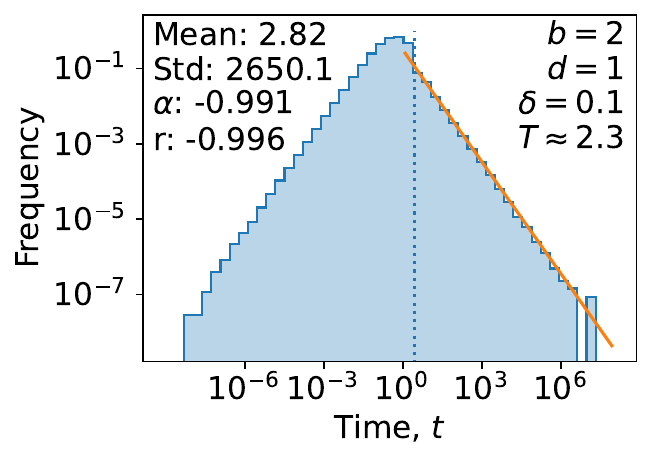}
\caption{\label{fig:distrib_H}
\textbf{Node depth distribution}. 
Sample of $10^8$ realisations of the random variable $H$ by the inversion of the cumulative probability function method. Deep nodes follow a power law distribution with parameter $\alpha=-1$ (orange line).}
\end{center}
\end{figure}

Proposition \ref{lem:coal_lineage} gives the cumulative probability function for the node depth $H$: $\mathbb P(H<t) = 1-\frac{1}{F(t)}$ (Figure \ref{fig:cdf_H}). 
Note that this function is defined piecewise for each cycle.

Because $F(t)\to \infty$ as $t\to\infty$, the cumulative distribution function of node depths tends to 1, which shows that $H$ cannot take the value $+\infty$, as is expected for critically (and also super-critically) branching populations.

The random variable $H$ can be easily sampled from its cumulative probability function. As illustrated in Figure \ref{fig:distrib_H},
the distribution of deep nodes (deeper than depth $1$) can be fitted by a power law with parameter $-1$ (criticality).

\subsection{Number of mutations}

To find the expected number of mutations within a droplet, the expected size of the full coalescent tree must be considered. 
This relies on the node depth distribution, conditioned to be lower than the duration of the experiment.
It results in the following:

\begin{figure}
\begin{center}
\includegraphics[width=0.5\textwidth]{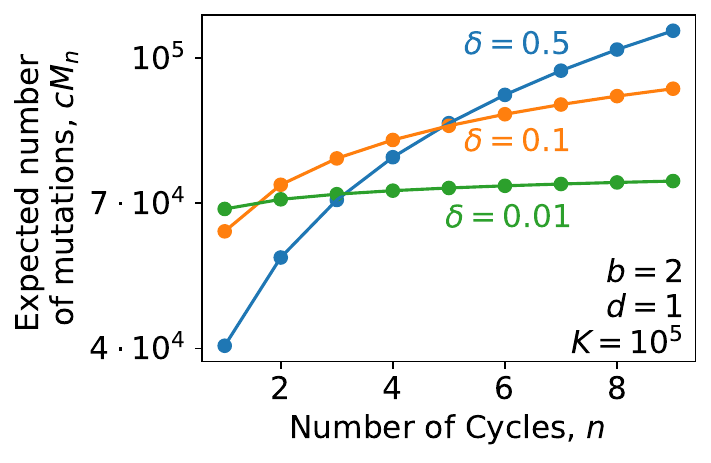}
\caption{\label{fig:mn_function_of_n}
\textbf{Expected number of mutations through experimental cycles}. 
The expected number of mutations increases with the number of experimental cycles. The initial number of cells is $c=K \delta$.
For one cycle ($n=1$), smaller dilution factors give better results (the curve $\delta = 0.5$ is lower than the curve for $\delta=0.001$). For more cycles, however, a larger dilution factor yields more mutations.}
\end{center}
\end{figure}

\begin{prop}[Number of mutations]
\label{lem:length_coal}
Let $M_n$ be the expected number of mutations (compared to the ancestral phenotype) accumulated in a lineage at the end of the $n$th cycle,
with dilution $\delta^*=e^{-(b-d)T}$, birth rate of cells $b$, death rate $d$, and mutation rate $\theta$.
\begin{equation}
    M_n = \theta s_n^* L_n,
\end{equation}
with $L_n$ the average length of the coalescent tree at the end of the $n$th cycle of an extant lineage started by one cell at $t=0$:
\begin{equation}
\label{eq:L_n}
L_n = F(nT) \int_0^{nT} \frac{dt}{F(t)},
\end{equation}
with $F$ the inverse tail distribution of the associated CPP. More specifically, when using the expression of $F$ from Proposition \ref{lem:coal_lineage}:
\begin{equation}
L_n = \left(1 + \frac br n (e^{rT}-1) \right) \sum_{k=0}^n \frac{rT - \log \left ( \frac{k (e^{rT} - 1) + \frac rb }{(k+1) (e^{rT} - 1) + \frac rb } \right ) }{ bk(e^{rT}-1)-d}
\end{equation}

\begin{flushright}
(Proof page \pageref{proof:length_coal}.)
\end{flushright}
\end{prop}

\begin{figure}
\begin{center}
\includegraphics[width=0.5\textwidth]{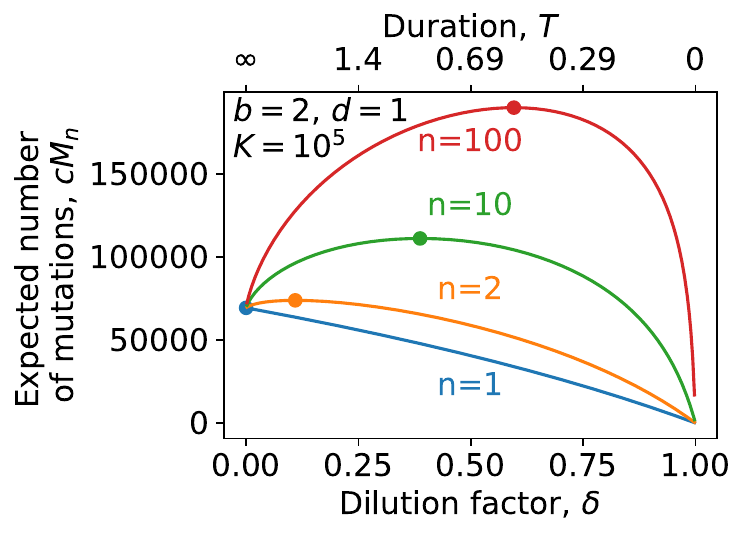}
\caption{\label{fig:mn_function_of_delta}
\textbf{Expected number of mutations for different dilution factors}. The initial number of cells is $c=K\delta$.
For a single cycle ($n=1$), smaller dilution factors always yield more mutations. 
However, if more than one cycle is performed ($n\ge 2$), there is a non-zero optimal dilution factor that maximises the expected number of mutations found in the population.}
\end{center}
\end{figure}

\begin{figure}
\begin{center}
\includegraphics[width=0.5\textwidth]{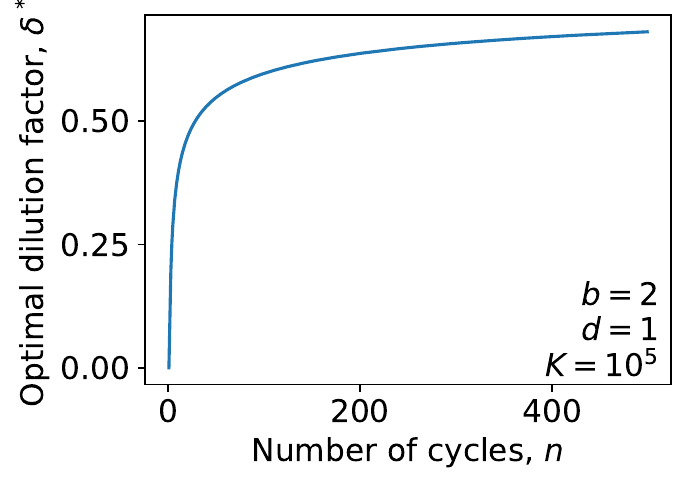}
\caption{\label{fig:opt_delta_function_of_n}
\textbf{Mutation-optimising bottleneck size as a function of the number of cycles}. The initial number of cells is $c=K\delta$. The bottleneck size that optimises the expected number of mutations is increasing with the number of cycles performed.}
\end{center}
\end{figure}

Proposition \ref{lem:length_coal} shows that the expected number of mutations $M_n$ accumulated in a lineage is, naturally, proportional to the mutation rate $\theta$. The expected number of mutations is also proportional to the expected coalescent tree length of a single extant lineage $L_n$ (weighted by the proportion of lineages that actually survive $s_n$).
This expected coalescent length is parameterised by the birth and death rates of the cells, but also the duration of the growth phase $T$.
The proportion of lineages that actually survives plays an important role here. 
Consider a massively parallel experiment with the goal of screening as many mutations as possible. 
To find the optimal protocol, one must not only seek to optimise the number of mutations that accumulate on one lineage $\theta L_n$ but also discount this number by the proportion of them that are extinct $s_n$.
Otherwise, one might waste a lot of resources conducting an experiment consisting mainly in serial transfer of droplets devoid of cells altogether. 
Finally, since the number of accumulated mutations is proportional to the number of initial lineages, increasing this number by producing more numerous or bigger droplets (with the same initial occupancy) proportionally increases the number of mutations.

The expected number of mutations increases indefinitely with the number of cycles (Figure \ref{fig:mn_function_of_n}). 
However, the rate of increase is tied to the dilution bottleneck and tends to slow down when the number of cycles increases.
Note that for a small number of cycles, the expected number of mutations increases with a higher dilution: one cycle with a dilution by two yields less diversity than one cycle with a dilution by one hundred.
However, if ten cycles are performed, a dilution by two yields more diversity. 
This illustrates a trade-off between a harsh bottleneck, which allows long cycles and thus potentially many mutations but leads to the loss of most extant mutations (due to founder effects), and a softer bottleneck that allows for fewer mutations to accumulate during each cycle but compounds more because fewer mutations are lost.

Figure \ref{fig:mn_function_of_delta} clarifies the link between the expected number of mutations and the dilution bottleneck. 
Note that smaller bottleneck sizes are compensated by longer cycles because the experiment is supposed to be performed in optimal conditions ($\delta=e^{-rT}$). 
If there is only one cycle ($n=1$), the maximal expected number of mutations is reached when the dilution bottleneck is vanishingly small ($\delta \to 0$) and the cycle length adequately long ($T\to \infty$). 
However, if there is more than one cycle ($n>1$), the expected number of mutations reaches a maximum value between $\delta=0$ and $\delta=1$.
This maximum-diversity dilution bottleneck value increases with the number of cycles (Figure \ref{fig:opt_delta_function_of_n}).
Thus, the dilution bottleneck should be adjusted to the expected duration of the experiment in terms of cycle number to maximise accumulation of neutral mutations.

Overall, the expected number of neutral mutations accumulated by the population increases through time and can be optimised by appropriately choosing a bottleneck size that optimises the trade-off between accumulating new mutations and not losing old ones.

Note that $L_n$ behaves like $Tn\ln(n)$ (\cite{lambert_allelic_2009}, Theorem 2.4); thus, the expected neutral diversity after $n$ cycles is equivalent to $\theta s_n T n\ln(n)$. 

Thanks to \eqref{eqn:survival-in-optimal-regime}, and because $\delta = \delta^\star = e^{-rT}$, we obtain:
\begin{equation*}
M_n \sim_{n \to + \infty}  \frac{\theta T r}{b(1-e^{-rT})}\ln(n).
\end{equation*}

The mere number of mutations contains little information about the diversity within a droplet.
Indeed, some of those mutations could be born by a single individual, while others might be shared by the whole population.
The next section addresses this problem by exploring the mutation frequency spectrum.

\subsection{Mutation Frequency Spectrum}

A more precise assessment of the neutral diversity structure involves distinguishing between rare mutations (carried by few individuals) and frequent mutations (widespread within the population). 
The mutation frequency spectrum presents the proportion of mutations that are carried by a given number of individuals. 
The expected mutation frequency spectrum of a CPP can be deduced from the law of node depths $H$ (\citep{lambert_allelic_2009}, Theorem 2.2).
Indeed, the number of mutations carried by $i$ individuals is proportional to the length of the coalescent tree subtending $i$ leaves (Figure \ref{fig:group_of_three}). As a result:

\begin{figure}
\begin{center}
\includegraphics[width=0.3\textwidth]{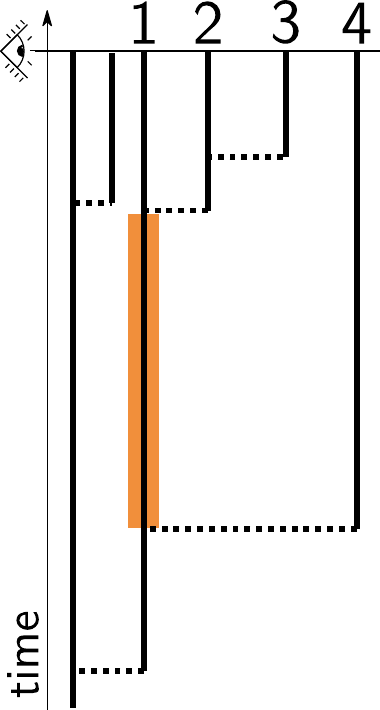}
\caption{\label{fig:group_of_three}
\textbf{Finding the mutation frequency spectrum}. 
The mutations shared by 3 individuals are the ones that arose within the orange region only. This region is delimited by $\max(H_{i+1}, H_{i+2})<t<\min(H_i, H_{i+3})$.
}\end{center}
\end{figure}

\begin{prop}[Mutation frequency spectrum]
\label{lem:mutatation_frequency_spectrum}
Consider the CPP $\mathcal T_n$, with overlaying mutations following a Poisson point process with intensity $\theta$.

Let $M^f_n$ be the expected number of mutations fixed in the population: that is, mutations shared by all individuals. 
Then:
\begin{equation}
    M^f_n = \theta s_n  \int_0^{nT} \frac{F(s)}{F(nT)} - \frac{F(nT)}{F(s)}  ds
\end{equation}

Let $a_u$ be the expected frequency of mutations that are shared by $u>0$ individuals in the limit of a large sample of the population.

\begin{equation}
   a_u = \theta \int_0^{nT} \left(1- \frac{\frac{1}{F(x)}-\frac{1}{F(nT)}}{1 - \frac{1}{F(nT)}} \right)^{u-1}
    \left ( \frac{\frac{1}{F(x)}-\frac{1}{F(nT)}}{1 - \frac{1}{F(nT)}}\right)^2 dx
\end{equation}

\begin{flushright}
(Proof page \pageref{proof:mutation_frequency_spectrum}.)
\end{flushright}
\end{prop}

As is expected for neutral diversity, Proposition \ref{lem:mutatation_frequency_spectrum} shows that the mutation frequency spectrum is proportional to the mutation rate $\theta$, meaning that an increasing proportion of individuals carry mutations if the rate increases, but that this does not change the relative frequency of the size of groups carrying a given mutation.

\begin{figure}
\begin{center}
\includegraphics[width=0.5\textwidth]{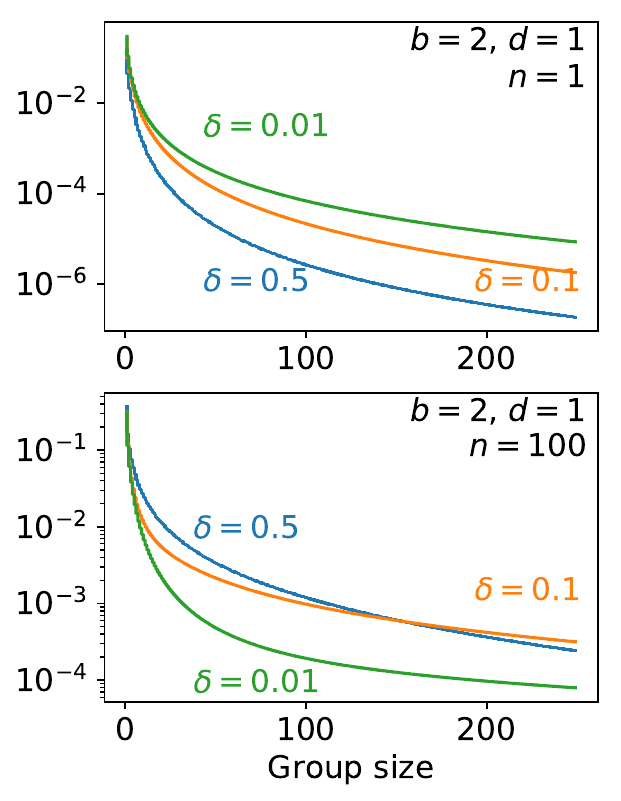}
\caption{\label{fig:mfs}
\textbf{Mutation frequency spectrum}. Give the frequency of mutations carried by a given number of individuals in a large sample of the population. \emph{Top:} After $n=1$ cycle. \emph{Bottom:} After $n=100$ cycles.}
\end{center}
\end{figure}

Figure \ref{fig:mfs} shows the mutation frequency spectra for one and for a hundred cycles, and for three different dilution factors.
Note that for a single cycle, harsher bottlenecks (i.e., smaller $\delta$ and correspondingly longer cycle duration $T$) increase the tail of the distribution (i.e., there are more mutations that are shared by many individuals). 
This effect of $\delta$ and $T$ is not as simple when considering several cycles. For $n=100$, the distribution is more heavy-tailed when the bottlenecks are soft ($\delta=0.5$) than when they are harsh ($\delta=0.001$). 

\begin{figure}
\begin{center}
\includegraphics[width=0.9\textwidth]{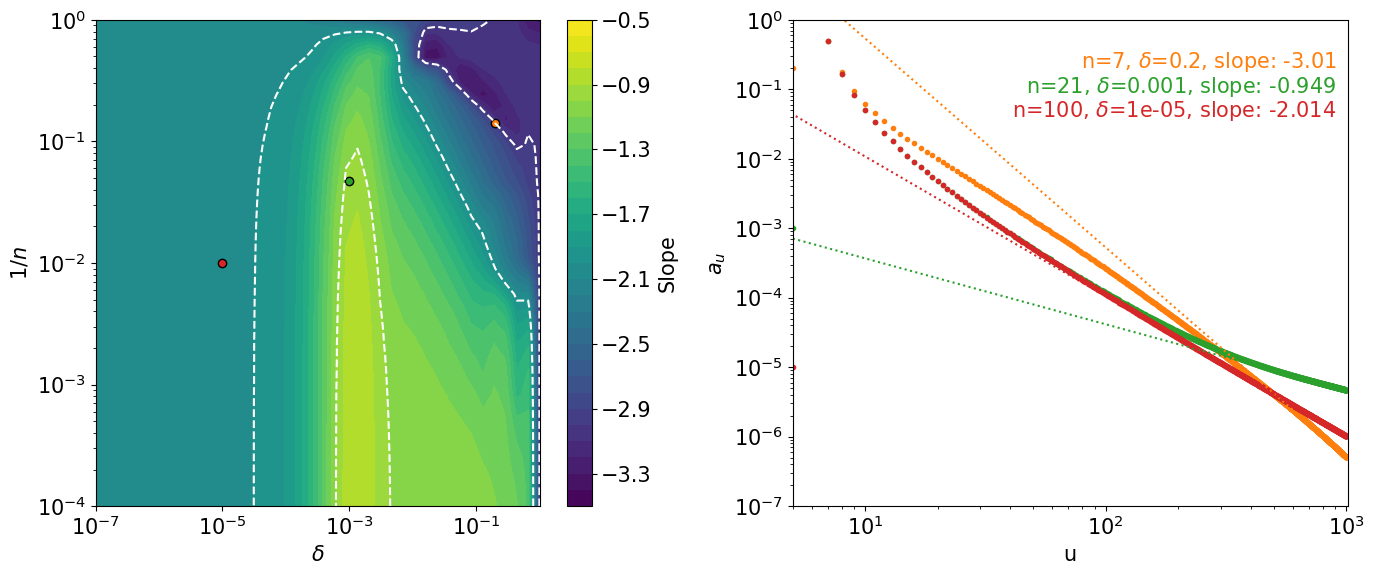}
\caption{\label{fig:mfs_tail}
\textbf{Mutation frequency spectrum power law tail}. Left: Slope of the regression $\alpha$, such that $log(a_u) = \alpha log(u) + b$, for $u>500$. For a branching process without bottleneck, the expected value is $-2$; it is $-1$ for a Moran process and $-3$ for a Coalescent Point Process with uniform node depths. White dashed lines are isolines $-1$, $-2$, $-3$. Right: three illustrative mutation frequency spectra with associated regression lines. The location of the three spectra in the parameter space is represented by colour-matching dots on the left heat-map. 
}
\end{center}
\end{figure}

Figure \ref{fig:mfs_tail} shows the power law tails of mutation frequency spectra.

When the coalescent tree is a Kingman coalescent, corresponding to a long-lived population with approximately constant size, the mutation frequency spectrum has power law with exponent $-1$ (harmonic spectrum, Ewens' sampling formula, \citep{ewens_sampling_1972}). In our setting, this happens when $\delta$ is close to 1 (constant population size) and $nT$ is large (long-lived population). For a fixed growth rate $r$, because $\delta e^{rT}=1$, this means that ($1-\delta$ is small but) $n(1-\delta)$ is large, as in the yellow region of the heat map of Figure \ref{fig:mfs_tail}.

When the coalescent tree is a Yule tree, corresponding to full, unbounded growth, the mutation frequency spectrum has power law with exponent $-2$ \citep{lambert_allelic_2009,dinh_statistical_2020,durrett_population_2013}. This is what happens for small $\delta$, regardless of $n$, as in the turquoise region of Figure \ref{fig:mfs_tail}. Indeed, when $\delta$ is small, $T$ is large (for fixed $r$), so that all derived mutations occurred since the last bottleneck.

A third case stands out in our setting when $n$ is not too large and $\delta$ is sufficiently close to 1 that very few births/coalescences occur over the time interval of length $nT$, which happens when $n(1-\delta)=O(1)$. In this case, conditioning the CPP to have coalescences smaller than $nT$ (an event of vanishing probability) yields a CPP with uniform node depths, which gives rise to a power law mutation frequency spectrum with exponent $-3$, as in the deep blue region of Figure \ref{fig:mfs_tail}.

\begin{figure}
\begin{center}
\includegraphics[width=0.5\textwidth]{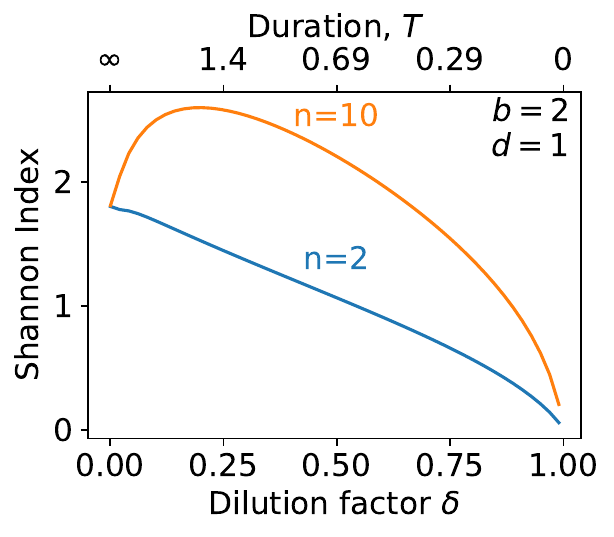}
\caption{\label{fig:shannon}
\textbf{Shannon Entropy of the mutation frequency spectrum}. 
Defined as $S = - \sum_i a_i \log(a_i)$}
\end{center}
\end{figure}

The information entropy of the mutation frequency spectrum can be used to systematically explore the effect of $\delta$ on the shape of the distribution. Figure \ref{fig:shannon} shows that if more than one cycle is performed, there is a value of $\delta$ that is expected to optimise the information entropy of the mutation frequency spectrum. This value is different from the value that optimises the number of mutations (Figure \ref{fig:mn_function_of_delta}).
Thus, there is a trade-off between accumulating many mutations and having a diverse mutation frequency spectrum. 
The decision to fix $\delta$ in order to optimise one or the other depends on the goal of the experiment.

To sum up, higher dilution factor or longer duration of droplet growth cycles result in longer trees and increased diversity, even though the population may risk going extinct. Extinction of the population marks the ``death'' of the culture and is the eventual fate of a serial transfer experiment in the limit of many cycles. 
In a nested populations' design, the cultures can also ``reproduce'' and replace the extinct ones. This has far-reaching consequences for the genealogy of the cells, as shown in the next section.

\section{Diversity in Dividing Droplets}

Nested populations' design differs from simple serial transfer in parallel cultures by the opportunity for the cultures (droplets, tubes or other compartments) to be subject to birth and death themselves. At each cycle, some cultures may be removed from the experiment, while others can be duplicated, usually by dispatching samples of the original droplet in several new fresh medium compartments (rather than one in regular parallel serial transfer experiments).

\begin{figure}
\begin{center}
\includegraphics[width=\textwidth]{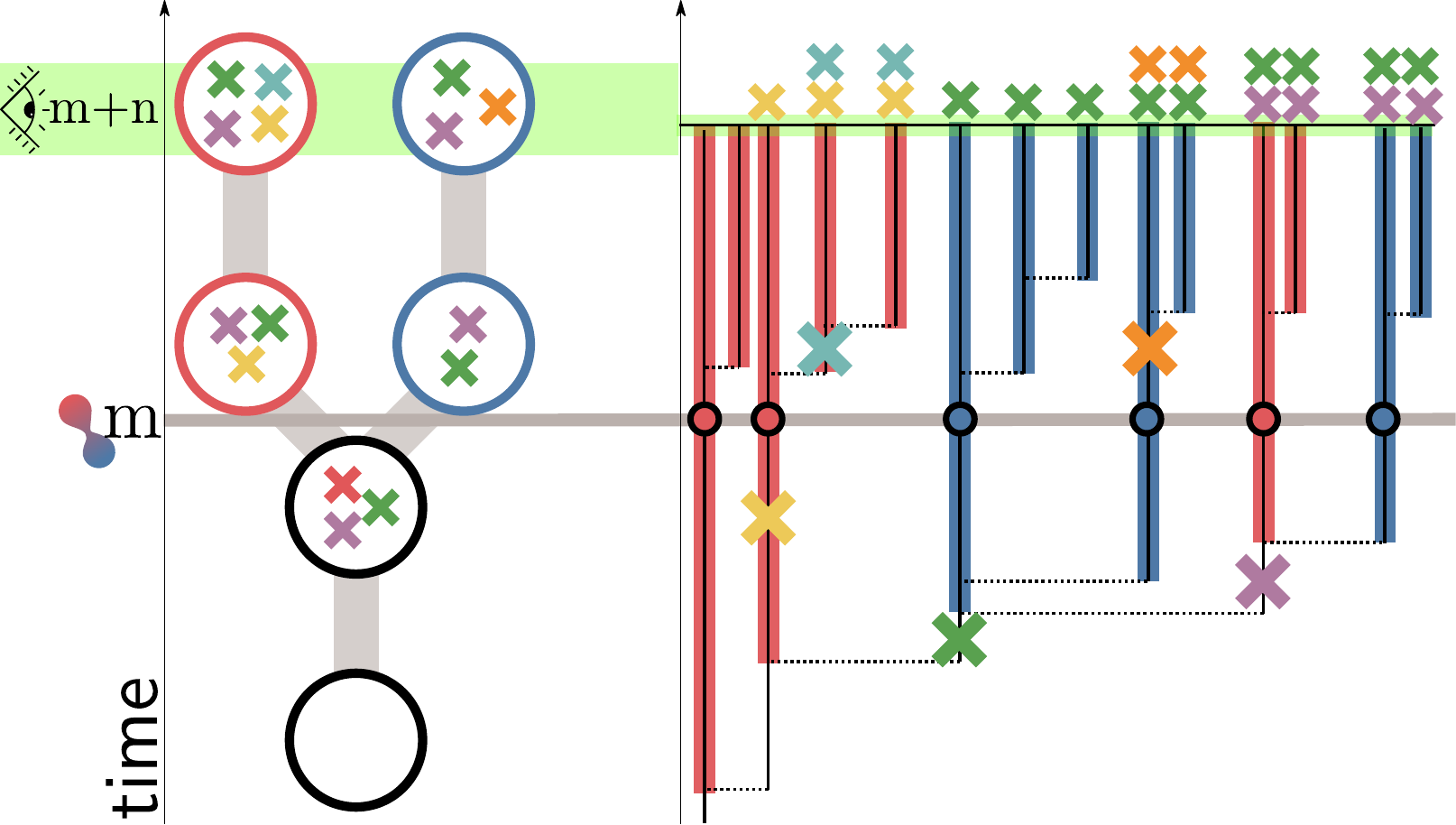}
\caption{\label{fig:nested_coal}
\textbf{Droplet and cell nested coalescent trees}. 
Left: coalescent tree of droplets. 
Right: coalescent tree of cells. 
An ancestral droplet lineage (black) is diluted into two offspring droplets (red, blue). 
At the time when the droplets are split (cycle $m$), the cell lineages within the ancestral droplets are assigned a colour (red or blue) that indicates the daughter droplet to which they are sent.  
Mutations appear along the genealogy of cells (crosses). 
Some mutations appear before the split (yellow, green, purple) and are found in both droplets (green, purple) if they are sampled by both droplet lineages, or within a single droplet (yellow) if they are segregated by the dilution.
Other mutations appear after the split (light blue, orange) and are only found in one of the droplet lineages.
}
\end{center}
\end{figure}

This section focuses on the consequences of imposing a droplet-level birth-death process on the neutral diversity.
To this end, consider the simple scenario (depicted in Figure \ref{fig:nested_coal}) of a pair of droplets that share a common ``droplet ancestor'' several cycles in the past.
The two droplets differ by the initial sampling performed in their common ancestor and also by all new mutations accumulated since they became isolated.

In the following, the cells follow a super-critical linear birth-death process with parameters $b-d=r>0$. 
The parameters of the population structure are supposed to be optimal in the sense of Section \ref{sec:optimal}: each cycle has a duration $T^*=-r^{-1}\ln{(cK^{-1}})$, and each lineage has an independent probability of being sampled at a bottleneck of $\delta^*=e^{-rT^*}$. Consider that the droplet split happens at cycle $m$ and the observation occurs at cycle $m+n$.

\subsection{Survival probability}

First, let us consider the probability that a lineage spawned by a single cell $n$ cycles in the past is not extinct within both droplets.
The key to establish this probability is to recognise that the lineage undergoes a bottleneck with survival probability $\delta$ at each cycle, except the cycle of the droplet split where each cell has a probability $2\delta$ to survive.
Indeed, two inoculation volumes are concurrently sampled from the ancestral droplet and dispatched into two offspring (Figure \ref{fig:dilution_process_kdrops}).
Thus:

\begin{figure}
\begin{center}
\includegraphics[width=0.5\textwidth]{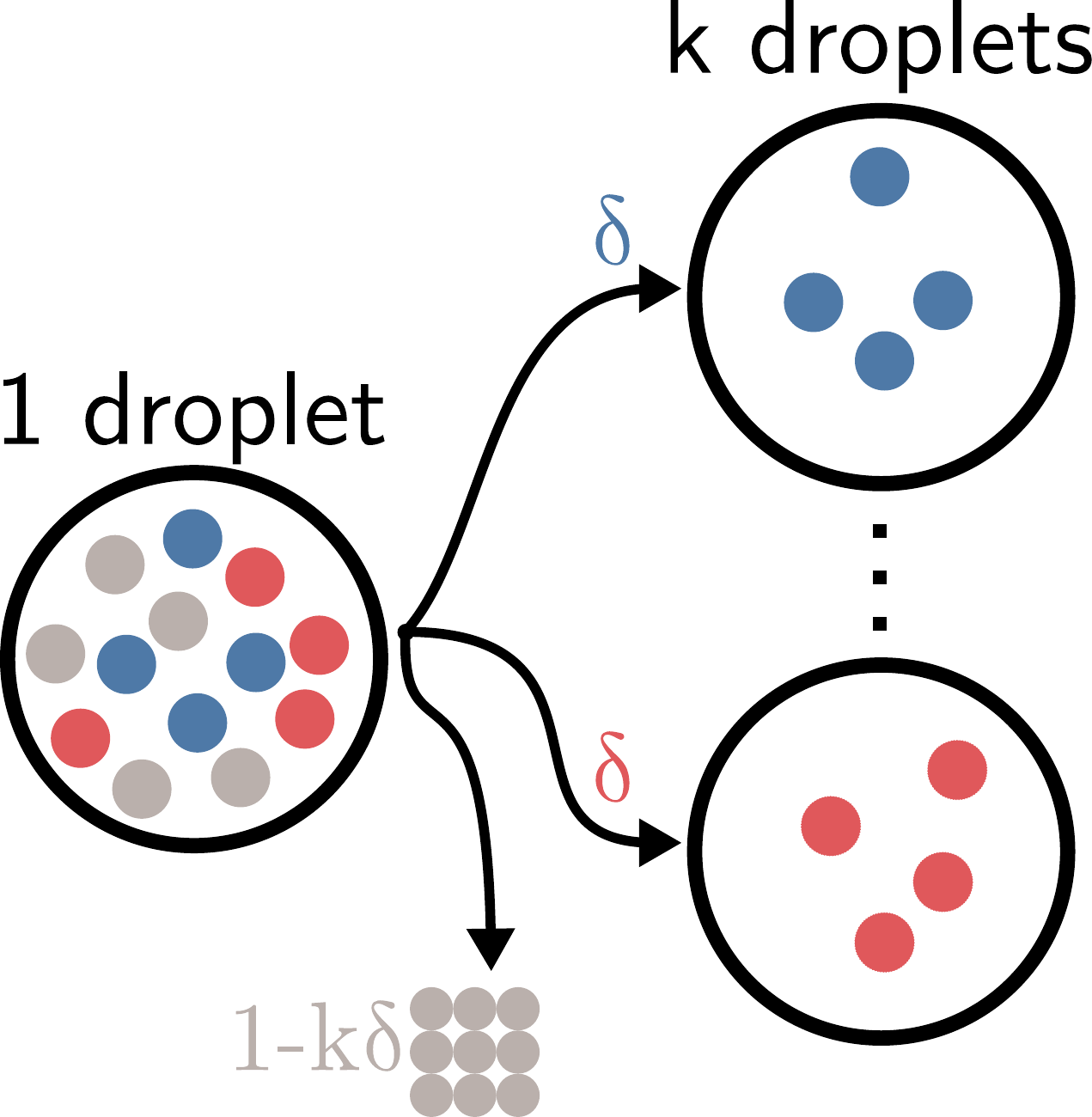}
\caption{\label{fig:dilution_process_kdrops}
\textbf{Droplet Splitting Process}. 
When a droplet is split into $k$ droplets, individual cells are independently selected to be transferred to the next cycle (with probability $\delta$ for each new droplet) or be discarded (with probability $1-k\delta$).}
\end{center}
\end{figure}

\begin{prop}[Survival probability - Split droplet]
\label{lem:survival2drops}

Cells within droplets in serial transfer are modelled by a linear birth-death process with constant parameters $b$ and $d$, subject to periodic bottlenecks $\delta$ every duration $T$ for $m+n$ cycles, except at cycle $m$ where $k$ droplets are produced (instead of one). Each new droplet is submitted independently to the serial transfer procedure for the remaining $n$ cycles.

The probability $s_{k,m,n}$ that a lineage spawned by a single cell at the beginning of the first cycle is not extinct at the end of the $(n+m)$-th cycle is given by:
\begin{equation}
s_{k,m,n} = 1 - h(Q_1 Q^{m-1}_\delta Q_{k\delta} Q^{n-1}_\delta, 0),
\end{equation}
where the matrix $Q_\varepsilon$ for $\varepsilon \in [0,1]$ and the function $h$ are defined in Proposition \ref{lem:survival_prob}.
\begin{flushright}
(Proof page \pageref{proof:survival2drops}.)
\end{flushright}
\end{prop}

Proposition \ref{lem:survival2drops} is similar in its conclusion to Proposition \ref{lem:survival_prob}, which treated the case of a simple serial transfer. 
However, the expression is considerably less easy to handle, as the iteration does not simplify into a single matrix power.

\subsection{Total diversity}

\begin{figure}
\begin{center}
\includegraphics[width=0.5\textwidth]{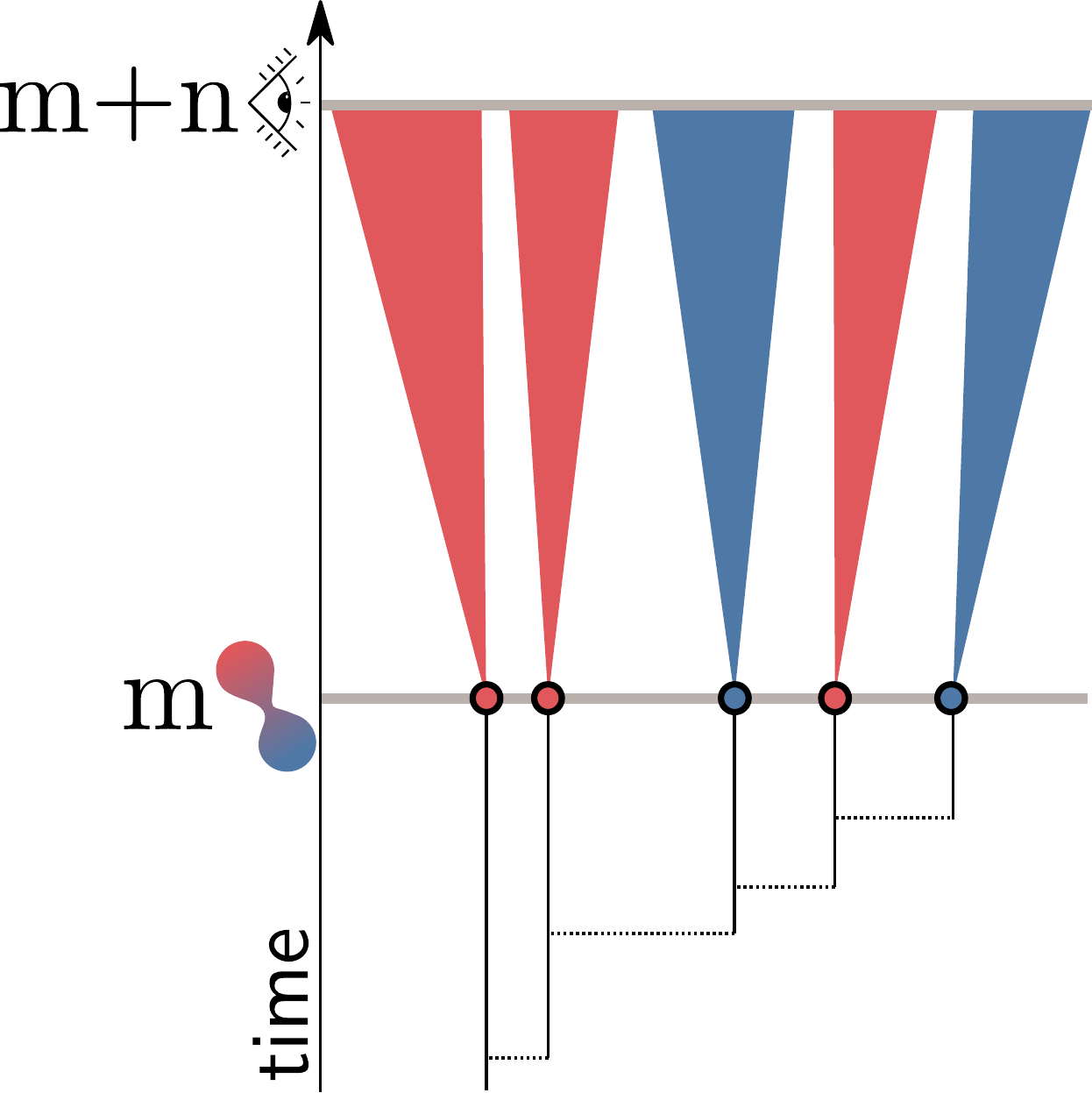}
\caption{\label{fig:redbluetotal}
\textbf{Finding the total diversity at cycle $n+m$ of droplets split at cycle $m$}.
All the lineages (triangles) spawned from the cells dispatched in one of the $k=2$ droplets (here red and blue) have the same expected length $L_{n}$.
The bottom part (or stump) of the tree (black) result from the sampling of a coalescent point process stopped at $mT$, with probability $\pi_{k,m,n}$, the probability that an extant lineage at cycle $m$ will be sampled in one of the two droplets and survive until cycle $n+m$.}
\end{center}
\end{figure}

As seen in Proposition \ref{lem:length_coal}, quantifying the total neutral diversity in an infinitely-many sites model is a matter of finding the total length of the coalescent tree (or forest) of the population.
Note that the full coalescent tree in Figure \ref{fig:redbluetotal} can be decomposed into a stump, before the splitting of droplets, and a corolla: another set of CPP (the corolla) sampled in one or the other droplet lineage.
As a result:

\begin{prop}[Total Diversity - Split droplet]
\label{lem:total_diversity2drops}
Let $M_{k,m,n}$ be the expected number of mutations accumulated in a lineage at cycle $n+m$ after the splitting of the initial droplet at cycle $m$ into $k=1,2\ldots \left \lfloor \frac{1}{\delta}\right \rfloor$ droplets. Then:
\begin{align*}
M_{k, m, n} =& \theta 
L_{k,m,n}\\
=& \theta s_m F^\dagger(mT) \left [\int_0^{mT} \frac{1}{F^\dagger(s)}ds + \int_0^{nT} \frac{F(nT)}{F(s)}ds \right ],
\end{align*}
where $L_{k,m,n}$ is the expected length of the coalescent tree of the population,
$F$ is the inverse tail distribution of the CPP with periodic bottlenecks, and $F^\dagger:= 1 -  k \delta s_{n} +  k \delta s_{n} F$,
is the inverse tail distribution of the same CPP submitted to sampling with probability $k\delta s_{n}$ at the present.

\begin{flushright}
(Proof page \pageref{proof:total_diversity2drops}.)
\end{flushright}
\end{prop}

\subsection{Private Diversity}

To assess the divergence between split droplets, one can compute the expected number of \emph{private} mutations (i.e., mutations that are only found in a single of the $k$ split droplets).
This number is the sum of all mutations that occur in the droplets after the splitting time $mT$ (i.e.,~mutations in the corolla), plus all the mutations that occur before the splitting time but in a lineage that only segregates in a single droplet (i.e.,~mutations in the stump). 
Since all the droplets are interchangeable, this value is identical for the $k$ droplets.
Overall, this number is proportional to the red (or blue) part of the CPP in figure \ref{fig:nested_coal}.

\begin{prop}[Private Mutations - Split Droplet]
\label{lem:private_mutations}

Let a single droplet be split into $k=1\ldots\left \lfloor \frac{1}{\delta} \right \rfloor$ at cycle $m$. 
Let $M'_{k,m,n}$ be the expected number of mutations that are private to any of the $k$ droplets when observed at cycle $n+m$.

\begin{align*}
    M'_{k,m,n} 
    &= \underbrace{\mathcal  S_{k,m,n}}_{\text{stump}}  +  \underbrace{\mathcal C_{k,m,n}}_{\text{corolla}} \\
   &= \theta s_m F^\dagger(mT) \left [ \int_0^{mT} \frac{k ds}{F^\dagger(s)(1+(k-1)F^\dagger(s))} 
   + \int_0^{nT} \frac{F(nT)}{F(s)}ds 
   \right ],
\end{align*}
where $F$ is the inverse tail distribution of the CPP with periodic bottlenecks, and $F^\dagger$ is the inverse tail distribution of the same CPP submitted to sampling with probability $k\delta s_{n}$ at the present.

\begin{flushright}
(Proof page \pageref{proof:private_mutations}.)
\end{flushright}
\end{prop}

\begin{figure}
\begin{center}
\includegraphics[width=0.5\textwidth]{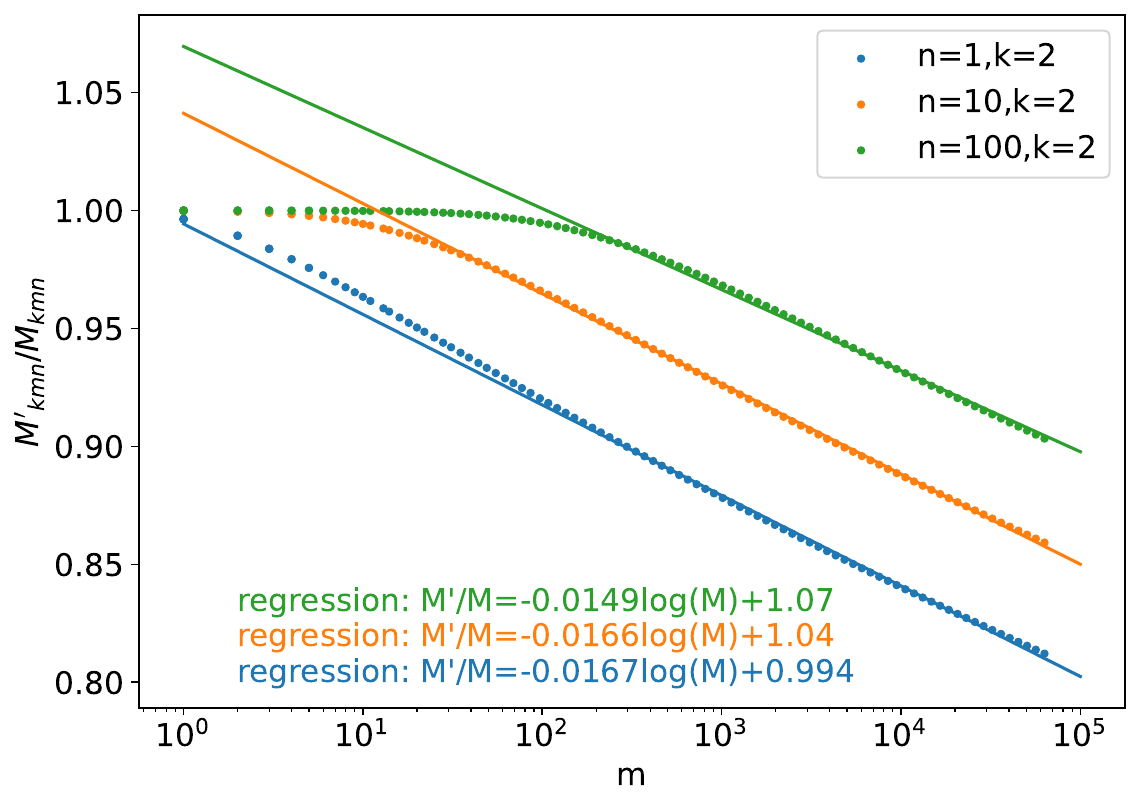}
\caption{\label{fig:mkmn_ratio}
\textbf{Expected proportion of private mutations in split droplets}. Droplets are grown for $m$ cycles, then split into $k=2$ new droplets and grown for $n$ new cycles. $b=1, d=0, \delta=e^{-rT}$}.
\end{center}\end{figure}

Figure \ref{fig:mkmn_ratio} shows the expected 
proportion of private mutations $M'_{k,m,n}/M_{k,m,n}$ in split droplets as a function of $m$, the number of cycles before splitting. If $m$ is low, there are no shared mutations among droplets and the ratio is close to $1$. If more cycles occur before the split, the proportion of private mutations decreases, tending toward 0 at a logarithmic rate.

Now, the main purpose of droplet splitting is to select and duplicate a phenotype of interest.
The last section explores, in the context of artificial selection, the advantage offered by a droplet-splitting process over the simple screening of parallel cultures in serial transfer.

\section{Artificial selection of droplets}

A practical application of a device that would allow the manipulation of small cultures of microbial organisms would be the artificial selection of phenotypes of interest. 
Suppose that a given phenotype of interest is reached after the accumulation of $\Theta$ mutations and that it is possible to detect the number of mutations fixed so far, by sequencing or direct observation of the cultures.

\begin{figure}[htbp]
\centering 
\includegraphics[width=.9\textwidth]{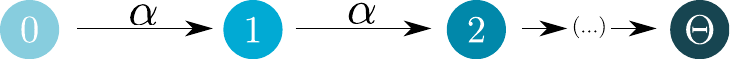}
\caption{
\textbf{Phenotypes.}
There are $\Theta+1$ possible phenotypes. 
A $j\text{-droplet}$ switches to the next phenotype $j+1$ at rate $\alpha$.}
\label{fig:stages}
\end{figure}

To formalise, let $D \in \mathbb  N^*$ be the number of droplets.
Each droplet $i$ is assigned a number $e_i=1,2,\ldots,\Theta$, corresponding to the number of fixed mutations.
Suppose that the time for a droplet to switch from $e_i=j$ to $e_i=j+1$ is exponentially distributed with parameter $\alpha = \frac{\rho}{ND}$, where $\rho$ is the mutation rate (that could be deduced from Proposition \ref{lem:mutatation_frequency_spectrum}), scaled by the number of droplets $D$ and the number of cycles $N$. We assume that the $D$ droplets are in state $0$ at time $0$. 
We also assume that the only possible transition is to accumulate a new mutation: that is, no reversion is possible, as illustrated in Figure \ref{fig:stages}.

To assess the advantage of droplet splitting, consider two scenarios, illustrated in Figure \ref{fig:propagation_mutations}:

\begin{enumerate}
\item \textbf{Without droplet selection} $D$ droplet lineages are started in state
$0$ at $t=0$ and undergo serial transfer independently of each other.
\item \textbf{With droplet selection} $D$ droplet lineages are started in state $0$
at $t=0$. Once a fixed mutant is detected in a lineage, all the other droplets are killed and this lineage is split in $D$ new lineages.
\end{enumerate}

Let $\Gamma$ (respectively, $\Gamma^*$) be the random variable encoding the first time for a lineage to get to the state $\Theta \in  \mathbb N$ in the scenario without droplet selection (respectively with droplet selection). To compare them, consider their respective cumulative distribution functions:

\begin{figure}
\begin{center}
\includegraphics[width=\textwidth]{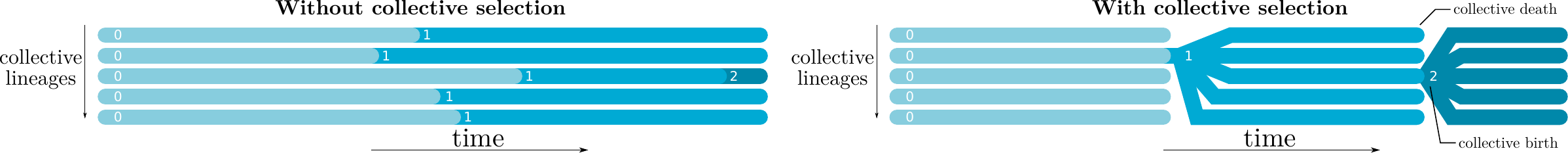}
\caption{\textbf{Propagation of mutations.} 
Without droplet selection, all the lineages accumulate mutations independently.
With droplet selection, once a mutation fixation is first detected, the droplet is split into $D$ lineages.}
\label{fig:propagation_mutations}
\end{center}
\end{figure}
  
\begin{prop}[Cumulative distribution functions]
\label{lem:timeToK}
The cumulative distribution function of $\Gamma^*$ is:

\begin{equation}
\mathbb P (\Gamma^* \leq x ) = 1 - e^{-x \rho} \left (\sum^{\Theta-1}_{u=0} \frac{( \rho x) ^{u}}{u!} \right )
\end{equation}

The cumulative distribution function of $\Gamma$ is:

\begin{equation}
\mathbb P (\Gamma \leq x ) =  1 - e^{-\rho x} \left ( \sum^{\Theta-1}_{u=0} \frac{(\rho x)^u}{D^u u!} \right )^D
\end{equation}

%When the number of mutational steps tends to infinity, the two cumulative distribution function are equivalent. However, for any finite number of mutational steps $\Theta$, 
The selective regime is always faster than the serial transfer regime, in the sense that $\Gamma^*$ is stochastically smaller than $\Gamma$ 
\begin{equation}
\mathbb P (\Gamma^* \leq x ) \geq \mathbb P (\Gamma \leq x ).
\end{equation}
In addition as the number of mutational steps $\Theta\to \infty$, the ratio
$\Gamma^*/\Gamma$ converges to $1/D$ almost surely.
\begin{flushright}
(Proof page \pageref{proof:timeToK}.)
\end{flushright}
\end{prop}

\begin{figure}
\begin{center}
\includegraphics[width=0.5\textwidth]{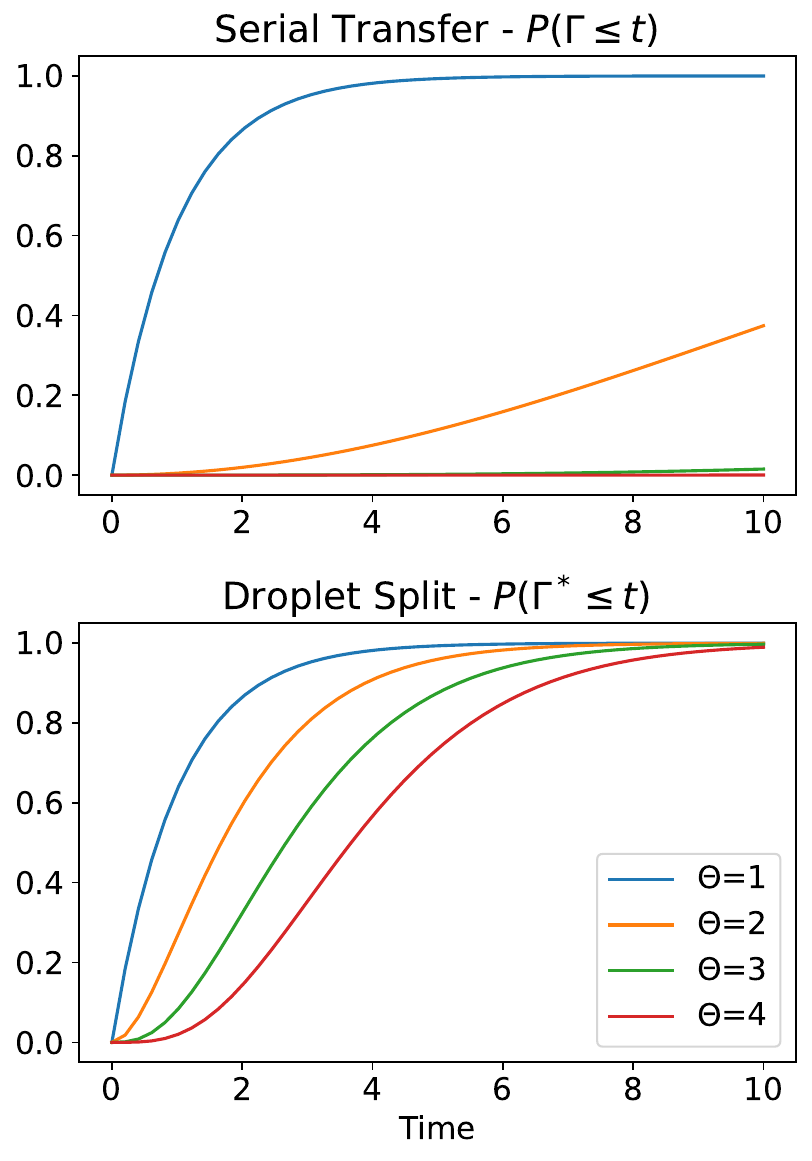}
\caption{\label{fig:cdf}
\textbf{Cumulative probability distribution} of the time to accumulate $\Theta$ mutations. With droplet splitting, accumulation of mutation is faster. $D=100$.
}\end{center}
\end{figure}

Proposition \ref{lem:timeToK} shows that droplet level selection---that is, the process of splitting a droplet in which an intermediate mutation was fixed---leads to reducing the time to reach the $\Theta\text{-th}$ mutation.
Figure \ref{fig:cdf} shows the shape of the cumulative probability function for both regimes, illustrating this advantage. 
This constitutes a simple use case for a device that allows the automated high-throughput manipulation of numerous cultures, such as the digital millifluidic analysers \citep{baraban_millifluidic_2011,boitard_growing_2015,cottinet_lineage_2016}. 

Note that this result is obtained by assuming that the detection of mutation is cost-free and error-free. A more advanced model of this system should tackle the problem of imperfect detection.

\section{Discussion}

This manuscript has laid the foundation for a theoretical understanding of the evolution of neutral diversity in massively parallel microbial evolution experiments with population splits. 
It was heavily inspired by ongoing engineering efforts to bring experimental evolution to digital millifluidics \citep{cottinet_diversite_2013,boitard_growing_2015,dupin_cultures_2018, doulcier_dropsignal_2019,ardre_leader_2022}.

In experimental microbiology, one desirable feature is to maximise the number of mutations accumulated within the cultures: for instance, to screen phenotypes of interest.
The result presented above showed that, in an optimal growth setting, where cells are growing with a constant birth and death rate, without density-dependence or competition, the population should be submitted to cycles whose duration is tailored to compensate the bottleneck imposed at each serial transfer. 
The choice of the bottleneck should be made according to the expected duration of the experiment: to optimise the expected number of mutations, small bottlenecks (i.e., killing most of the lineages) should be used when the number of cycles is small, while larger bottlenecks (i.e., lower dilution factor) should be used for long-term experiments.
Additionally, the expected number of mutations increases linearly with increasing droplet volume and with increasing number of droplets, which is a matter of technological progress as automation and larger droplet sizes are under consideration \citep{dupin_cultures_2018,postek_microfluidics_2022}. 
The mutation rate also increases linearly the number of expected mutations and can be manipulated by choosing mutator lineages or adding mutating chemicals to the culture broth. 
However, the potentially deleterious effects of this method might prevent using it in practical cases.

In long-term evolution experiments \citep{kawecki_experimental_2012,van_den_bergh_experimental_2018}, serial transfer is imposed by the need to replenish nutrients available to the cells. 
It is possible to build devices ensuring that a continuous flow of nutrient washes over the culture (for large volumes, see chemostats, or morbidostats \citep{toprak_building_2013}; in microfluidics, see mother machines \citep{potvin-trottier_microfluidics_2018}.
However, these methods are usually more prone to contamination. 
In contrast, periodically diluting the culture in a fresh medium is simple and robust.

The nested populations' design differs from traditional serial transfer of parallel cultures because it allows a birth-death process at the level of droplets. 
Serial transfer is pervasive in experimental evolution \citep{kawecki_experimental_2012} and has received extensive theoretical treatment.
This article focused solely on neutral diversity, considering that mutations have no effect on the demography of the cells.
However, the effect of beneficial mutations has been studied in serial transfer settings: 
in particular, the probability of losing a beneficial mutation due to repeated bottlenecks \citep{wahl_models_2000,wahl_evaluating_2002,wahl_probability_2001,wahl_survival_2015} and the effect of bottlenecks on the evolutionary path when multiple beneficial mutations exist \citep{gamblin_bottlenecks_2023}.
These results should be extended to the nested populations' design in the future.

The ``death'' of a droplet in our framework can have different origins. 
First is the experimentalist actively discarding some droplets with undesired characteristics. 
Second, it can arise from the fact that some cultures are effectively empty due to high dilutions in the previous cycle and may be replaced in the next cycle by cells from a non-empty culture. 
Finally, it can also be a consequence of the experimental protocol. 
Milli- and microfluidics compartments are usually produced in large numbers, while measurements are performed on all compartments. The retrieval of all the compartments' contents might not be practically possible or even desirable when they are too numerous. 
In all these cases, some droplets will not be used to generate the population of the next cycle, and some of them will have to be split if the droplet population is to keep a constant size.  

The use of a non-saturating population dynamics in this manuscript is a simplification that should be carefully taken into account when transposing the results of this work to the design of experiments. 
Nonetheless, if the cycle duration is short enough that the population is stopped during the exponential phase, the heuristics developed in this manuscript should hold. 
There are, however, two phenomena that were not modelled here and that will probably muddy the neutral pattern that was described. 
First is the absence of mutations affecting birth and death rates. 
If most point mutations can be safely considered neutral, some rare mutations can affect the ability of the cells to reproduce. 
If the mutations are beneficial, they will increase in proportion within the population and will change the relative frequency of all neutral mutations, by favouring the ones carried by the same strand of DNA (a phenomenon known as ``hitch-hiking'' \citealt{fay_hitchhiking_2000}). 
Second, horizontal gene transfer might allow the uncoupling of the mutation transmission from the genealogy \citep{dutta_horizontal_2002}, muddying the pattern even further.
In general, complete models should combine neutral and non-neutral approaches to describe microbial communities \citep{ofiteru_combined_2010,dumbrell_relative_2010}.

The nested populations' design also differs from trait groups' \citep{wilson_theory_1975} or transient compartments' \citep{blokhuis_selection_2018} population structure because migration between compartments is prevented.
As a consequence, it is possible to construct a non-ambiguous genealogy of the cultures. In practice, serial transfer design offers a natural way to implement the birth-death process by diluting some cultures into several new compartments (the droplet splitting) and discarding others.

Finally, this manuscript touches briefly on the problem of artificial selection using a nested populations' design. 
This was done by considering the accumulation of neutral mutations. 
A more complete model of artificial selection would, however, take into account interactions between individuals and potentially the selection of whole communities. 
Community-level selection has been the subject of both experimental \citep{swenson_artificial_2000,panke-buisse_selection_2015} and theoretical inquiries \citep{mueller_engineering_2015,arias-sanchez_artificially_2019,xie_simulations_2019,doulcier_eco-evolutionary_2020}.

Overall, the results presented in this manuscript should be considered as a way to build intuition about the experimental system while providing a null-model for diversity that could be compared to the actual patterns.
Inevitable differences have to appear, but the point of comparison that is offered by neutral evolution will allow a better description of the observed diversity.
Focusing on the part of the patterns that differ from this naive theoretical prediction will surely be fruitful---by showing that other mechanisms besides drift must be invoked.

\appendix
\section*{Reference}
%In Sections 2 and 3, the initial population contains initially $Z_0 = c$ 
%cells, and is submitted to a bottleneck with survival probability $\delta$ at the beginning of every cycle except the first one, meaning that bottlenecks occur at times $T,2T \ldots nT$. The $n$th cycle corresponds to the slice of time $[(n-1)T,nT)$. Thus "the end of cycle n" corresponds to the moment $nT^{-}$, just before the $n$th dilution. To illustrate: at the end of the third cycle, $t=3T^{-}$, and the population has experienced two bottlenecks at time $T$ and $2T$. 

%In section 4, the droplet is split into $k$ droplets at cycle $m$, meaning that the $m$th cycle starts with a bottleneck $k\delta$ instead of $\delta$. 

\begin{table}[!ht]
	\caption{Global Symbols reference}
 	\centering
\begin{tabular}{c|p{12cm}|p{3cm}}
\textbf{Symbol} & \textbf{Name} & \textbf{Reference}\\
\hline
$D$ & Number of droplets & \\ 
$T$ & Cycle duration & \\
$n$ & Number of cycles &\\
$K$ & Droplet carrying capacity &\\
$c$ & Initial number of particles &\\
$b,d,r$ & Particle birth rate, death rate and Malthusian parameter ($r:=b-d$) & \\
$\delta$ & Dilution factor (i.e., particle survival probability at a bottleneck) & \\
$\theta$ & Particle mutation rate & \\
$s_n$ & Probability that a lineage spawned by a single particle is not extinct at the end of the $n$th cycle. (At time $nT^-$, just before the $n$th dilution). $s_n=1-h_{Q^{n-1}R}(0)$ & Proposition \ref{lem:survival_prob}\\
$s_\infty$ & Limit probability that a lineage spawned by a single particle is not extinct after a large number of cycles. $s_\infty=\lim_{n\to\infty}s_n$. & Proposition \ref{lem:survival_prob_long_term}\\
$T^*$& Maximal cycle duration before reaching saturation. $T^*= r^{-1}\ln(cK^{-1})$ & Proposition \ref{lem:tstar}\\
$\delta^*$& Optimal dilution factor $\delta^*=e^{-rT}$ & Proposition \ref{lem:dstar}\\
$s^*_n$& Survival probability of a lineage spawned by a single cell before the $n$th dilution in the optimal regime in which $\delta=e^{-rT}$. & Proposition \ref{lem:snstar}\\
$\tilde F$ & Inverse tail distribution of the CPP without bottlenecks  & Proposition \ref{lem:coal_lineage} \\
$F$ & Inverse tail distribution of the CPP with bottlenecks. $F(t)=[\mathbb P(H>t)]^{-1}$ & Equation \ref{eq:F} and Proposition \ref{lem:coal_lineage} \\
$\tau_n$ & Coalescent tree of an extant lineage at the end of the $n$th cycle. It is a Coalescent Point Process with inverse tail distribution $F$ stopped at the first branch length larger than $nT$.& Proposition \ref{lem:length_coal}\\
$N(\tau_n)$& number of leaves of the coalescent tree $\tau_n$. Geometric random variable with expected value $F(nT)$. & Eq. \ref{eq:probanumberofleaves} and \ref{eq:expected_leaves_cpp}.\\
$L_n$ & Expected length of the coalescent tree $\tau_n$. $L_n=F(nT)\int_0^{nT^-}F(x)^{-1}dx$&  Proposition \ref{lem:length_coal} \\
$M_n$ & Expected number of mutations in a lineage after $n$ cycles. $M_n=\theta s_nL_n$&  Proposition \ref{lem:length_coal}  \\
$M^f_n$ & Expected number of fixed mutations in a lineage after $n$ cycles. &  
Proposition \ref{lem:mutatation_frequency_spectrum}\\
$M^s_n$ & Expected number of segregating mutations in a lineage after $n$ cycles. $M^s_n=M_n-M^f_n$ &  
Proposition \ref{lem:mutatation_frequency_spectrum}\\
$a_u$ & Expected frequency of mutations shared by $u>0$ individuals after $n$ cycles. &  
Proposition \ref{lem:mutatation_frequency_spectrum} \\
\end{tabular}
\end{table}
\begin{table}[!ht]
	\caption{Global Symbols reference (cont)}
 	\centering
\begin{tabular}{c|p{12cm}|p{3cm}}
\textbf{Symbol} & \textbf{Name} & \textbf{Reference}\\
\hline
$s_{k,m,n}$ & Survival probability at the end of the $(m+n)$th cycle of a lineage spawned by a single cell in a single droplet at the first cycle, that is was split into $k$ droplets at cycle $m$. & Proposition \ref{lem:survival2drops}\\
$\tau_{k,m,n}$ & Coalescent tree at the end of the $(m+n)$th cycle of a lineage spawned by a single cell in a single droplet at the first cycle, that is was split into $k$ droplets at cycle $m$. & Proposition  \ref{lem:total_diversity2drops}\\
$\pi_{k,n}$ & Probability that a lineage extant at the end of cycle $m$ just before the droplet is split into $k$ droplets will be extant at cycle $m+n$. $\pi_{k,n}=k \delta s_n$.\\
$F^\dagger$ & Inverse tail distribution of the stump tree. $F^\dagger(t) := F_{\pi_{k,n}}(t)= 1 - \pi_{k,n} + \pi_{k,n}F(t), t \in [0,mT]$&\\ 
$L_{k,m,n}$ & Expected length of the coalescent tree $\tau_{k,m,n}$. & Proposition \ref{lem:total_diversity2drops} \\
$M_{k,m,n}$ & Expected number of mutations accumulated at the end of the $(m+n)$th cycle of a lineage spawned by a single cell in a single droplet at the first cycle, that is was split into $k$ droplets at cycle $m$. $M=\theta s_{k,m,n}L_{k,m,n}$& \\
%$L^\dagger_{k,m,n}$ & Expected length of the stump tree at the end of the $m+n$th cycle of a lineage spawned by a single cell in a single droplet at the first cycle, that is was split into $k$ droplets at cycle $m$. It is the part of the coalescent tree that lies before the $m$th cycle.  & Proposition \ref{lem:private_mutations}  \\

%$\psi_{k,m,n}$ & Number of mutations occurring in the stump tree that are only carried by cells that are found in a single droplet. & Proposition \ref{lem:private_mutations} \\
$M'_{k,m,n}$ & Expected number of mutations that are only found in a single droplet at the end of the $(m+n)$th cycle of a lineage spawned by a single cell in a single droplet at the first cycle, that is was split into $k$ droplets at cycle $m$. & Proposition \ref{lem:private_mutations} \\
\end{tabular}
\end{table}
\begin{table}[!ht]
	\caption{Global Symbols reference (cont.)}
 	\centering
\begin{tabular}{c|p{12cm}|p{3cm}}
\textbf{Symbol} & \textbf{Name} & \textbf{Reference}\\
\hline
$\alpha$ & Rate at which a lineage accumulate mutations $\alpha=\frac{\rho}{ND}$&  Proposition \ref{lem:timeToK} \\
$\Theta$ & Number of mutations to accumulate &  Proposition \ref{lem:timeToK} \\
$\Gamma$ & First time a lineage has accumulated $\Theta$ mutations without droplet selection &  Proposition \ref{lem:timeToK} \\
$\Gamma^*$ & First time a lineage has accumulated $\Theta$ mutations with droplet selection & Proposition \ref{lem:timeToK} \\
\end{tabular}
\end{table}

\bibliographystyle{apalike}
\bibliography{references}  

\begin{thebibliography}{}

\bibitem[Ardré et~al., 2022]{ardre_leader_2022}
Ardré, M., Doulcier, G., Brenner, N., and Rainey, P.~B. (2022).
\newblock A leader cell triggers end of lag phase in populations of
  {Pseudomonas} fluorescens.
\newblock {\em microLife}, 3:uqac022.

\bibitem[Arias-Sánchez et~al., 2019]{arias-sanchez_artificially_2019}
Arias-Sánchez, F.~I., Vessman, B., and Mitri, S. (2019).
\newblock Artificially selecting microbial communities: {If} we can breed dogs,
  why not microbiomes?
\newblock {\em PLOS Biology}, 17(8):e3000356.

\bibitem[Athreya and Ney, 1972]{athreya_branching_1972}
Athreya, K.~B. and Ney, P.~E. (1972).
\newblock {\em Branching {Processes}}.
\newblock Springer Berlin Heidelberg, Berlin, Heidelberg.

\bibitem[Baraban et~al., 2011]{baraban_millifluidic_2011}
Baraban, L., Bertholle, F., Salverda, M. L.~M., Bremond, N., Panizza, P.,
  Baudry, J., Visser, J. A. G. M.~d., and Bibette, J. (2011).
\newblock Millifluidic droplet analyser for microbiology.
\newblock {\em Lab on a Chip}, 11(23):4057--4062.

\bibitem[Barton et~al., 2002]{barton_neutral_2002}
Barton, N.~H., Depaulis, F., and Etheridge, A.~M. (2002).
\newblock Neutral {Evolution} in {Spatially} {Continuous} {Populations}.
\newblock {\em Theoretical Population Biology}, 61(1):31--48.

\bibitem[Barton et~al., 2013]{barton_modelling_2013}
Barton, N.~H., Etheridge, A.~M., and Véber, A. (2013).
\newblock Modelling evolution in a spatial continuum.
\newblock {\em Journal of Statistical Mechanics: Theory and Experiment},
  2013(01):P01002.

\bibitem[Black et~al., 2020]{black_ecological_2020}
Black, A.~J., Bourrat, P., and Rainey, P.~B. (2020).
\newblock Ecological scaffolding and the evolution of individuality.
\newblock {\em Nature Ecology \& Evolution}, 4(3):426--436.

\bibitem[Blokhuis et~al., 2018]{blokhuis_selection_2018}
Blokhuis, A., Lacoste, D., Nghe, P., and Peliti, L. (2018).
\newblock Selection {Dynamics} in {Transient} {Compartmentalization}.
\newblock {\em Physical Review Letters}, 120(15):158101.

\bibitem[Boitard et~al., 2015]{boitard_growing_2015}
Boitard, L., Cottinet, D., Bremond, N., Baudry, J., and Bibette, J. (2015).
\newblock Growing microbes in millifluidic droplets.
\newblock {\em Engineering in Life Sciences}, 15(3):318--326.

\bibitem[Cottinet, 2013]{cottinet_diversite_2013}
Cottinet, D. (2013).
\newblock {\em Diversité phénotypique et adaptation chez {Escherichia} {Coli}
  etudiéees en millifluidique digitale}.
\newblock PhD thesis, Université Pierre et Marie Curie.

\bibitem[Cottinet et~al., 2016]{cottinet_lineage_2016}
Cottinet, D., Condamine, F., Bremond, N., Griffiths, A.~D., Rainey, P.~B.,
  Visser, J. A. G. M.~d., Baudry, J., and Bibette, J. (2016).
\newblock Lineage tracking for probing heritable phenotypes at single-cell
  resolution.
\newblock {\em PLOS ONE}, 11(4):e0152395.

\bibitem[Delgado-Baquerizo et~al., 2016]{delgado-baquerizo_lack_2016}
Delgado-Baquerizo, M., Giaramida, L., Reich, P.~B., Khachane, A.~N., Hamonts,
  K., Edwards, C., Lawton, L.~A., and Singh, B.~K. (2016).
\newblock Lack of functional redundancy in the relationship between microbial
  diversity and ecosystem functioning.
\newblock {\em Journal of Ecology}, 104(4):936--946.
\newblock \_eprint:
  https://onlinelibrary.wiley.com/doi/pdf/10.1111/1365-2745.12585.

\bibitem[Dinh et~al., 2020]{dinh_statistical_2020}
Dinh, K.~N., Jaksik, R., Kimmel, M., Lambert, A., and Tavaré, S. (2020).
\newblock Statistical {Inference} for the {Evolutionary} {History} of {Cancer}
  {Genomes}.
\newblock {\em Statistical Science}, 35(1):129--144.

\bibitem[Doulcier, 2019]{doulcier_dropsignal_2019}
Doulcier, G. (2019).
\newblock Dropsignal - {Millifluidic} droplet trains analysis.
\newblock {\em Zenodo}.

\bibitem[Doulcier et~al., 2020]{doulcier_eco-evolutionary_2020}
Doulcier, G., Lambert, A., De~Monte, S., and Rainey, P.~B. (2020).
\newblock Eco-evolutionary dynamics of nested {Darwinian} populations and the
  emergence of community-level heredity.
\newblock {\em eLife}, 9:e53433.

\bibitem[Dumbrell et~al., 2010]{dumbrell_relative_2010}
Dumbrell, A.~J., Nelson, M., Helgason, T., Dytham, C., and Fitter, A.~H.
  (2010).
\newblock Relative roles of niche and neutral processes in structuring a soil
  microbial community.
\newblock {\em The ISME Journal}, 4(3):337--345.
\newblock Number: 3 Publisher: Nature Publishing Group.

\bibitem[Dupin, 2018]{dupin_cultures_2018}
Dupin, J.-B. (2018).
\newblock {\em Cultures multi-parallélisées en millifluidique digitale :
  diversité et sélection artificielle}.
\newblock PhD thesis, Université Pierre et Marie Curie.

\bibitem[Durrett, 2013]{durrett_population_2013}
Durrett, R. (2013).
\newblock Population genetics of neutral mutations in exponentially growing
  cancer cell populations.
\newblock {\em The Annals of Applied Probability}, 23(1):230--250.

\bibitem[Dutta and Pan, 2002]{dutta_horizontal_2002}
Dutta, C. and Pan, A. (2002).
\newblock Horizontal gene transfer and bacterial diversity.
\newblock {\em Journal of Biosciences}, 27(1):27--33.

\bibitem[Etheridge, 2008]{etheridge_drift_2008}
Etheridge, A.~M. (2008).
\newblock Drift, draft and structure: some mathematical models of evolution.
\newblock {\em Banach Center Publications}, 80:121--144.

\bibitem[Ewens, 1972]{ewens_sampling_1972}
Ewens, W.~J. (1972).
\newblock The sampling theory of selectively neutral alleles.
\newblock {\em Theoretical population biology}, 3(1):87--112.

\bibitem[Fay and Wu, 2000]{fay_hitchhiking_2000}
Fay, J.~C. and Wu, C.-I. (2000).
\newblock Hitchhiking {Under} {Positive} {Darwinian} {Selection}.
\newblock {\em Genetics}, 155(3):1405--1413.

\bibitem[Fuhrman, 2009]{fuhrman_microbial_2009}
Fuhrman, J.~A. (2009).
\newblock Microbial community structure and its functional implications.
\newblock {\em Nature}, 459(7244):193--199.
\newblock Number: 7244 Publisher: Nature Publishing Group.

\bibitem[Gamblin et~al., 2023]{gamblin_bottlenecks_2023}
Gamblin, J., Gandon, S., Blanquart, F., and Lambert, A. (2023).
\newblock Bottlenecks can constrain and channel evolutionary paths.
\newblock {\em Genetics}, 224(2):iyad001.

\bibitem[Hammerschmidt et~al., 2014]{hammerschmidt_life_2014}
Hammerschmidt, K., Rose, C.~J., Kerr, B., and Rainey, P.~B. (2014).
\newblock Life cycles, fitness decoupling and the evolution of
  multicellularity.
\newblock {\em Nature}, 515(7525):75--79.

\bibitem[Kassen et~al., 2000]{kassen_diversity_2000}
Kassen, R., Buckling, A., Bell, G., and Rainey, P.~B. (2000).
\newblock Diversity peaks at intermediate productivity in a laboratory
  microcosm.
\newblock {\em Nature}, 406(6795):508--512.
\newblock Number: 6795 Publisher: Nature Publishing Group.

\bibitem[Kawecki et~al., 2012]{kawecki_experimental_2012}
Kawecki, T.~J., Lenski, R.~E., Ebert, D., Hollis, B., Olivieri, I., and
  Whitlock, M.~C. (2012).
\newblock Experimental evolution.
\newblock {\em Trends in ecology \& evolution}, 27(10):547--560.

\bibitem[Lambert, 2009]{lambert_allelic_2009}
Lambert, A. (2009).
\newblock The allelic partition for coalescent point processes.
\newblock {\em Markov Processes And Related Fields}, 15(3):359--386.

\bibitem[Lambert and Stadler, 2013]{lambert_birthdeath_2013}
Lambert, A. and Stadler, T. (2013).
\newblock Birth–death models and coalescent point processes: {The} shape and
  probability of reconstructed phylogenies.
\newblock {\em Theoretical Population Biology}, 90:113--128.

\bibitem[Lewontin, 1970]{lewontin_units_1970}
Lewontin, R.~C. (1970).
\newblock The {Units} of {Selection}.
\newblock {\em Annual Review of Ecology and Systematics}, 1:1--18.
\newblock tex.ids: lewontinUnitsSelection1970a.

\bibitem[Li et~al., 2022]{li_design_2022}
Li, X., Zhou, Z., Li, W., Yan, Y., Shen, X., Wang, J., Sun, X., and Yuan, Q.
  (2022).
\newblock Design of stable and self-regulated microbial consortia for chemical
  synthesis.
\newblock {\em Nature Communications}, 13(1):1554.
\newblock Number: 1 Publisher: Nature Publishing Group.

\bibitem[Maron et~al., 2018]{maron_high_2018}
Maron, P.-A., Sarr, A., Kaisermann, A., Lévêque, J., Mathieu, O., Guigue, J.,
  Karimi, B., Bernard, L., Dequiedt, S., Terrat, S., Chabbi, A., and Ranjard,
  L. (2018).
\newblock High {Microbial} {Diversity} {Promotes} {Soil} {Ecosystem}
  {Functioning}.
\newblock {\em Applied and Environmental Microbiology}, 84(9):e02738--17.
\newblock Publisher: American Society for Microbiology.

\bibitem[Milshteyn et~al., 2014]{milshteyn_mining_2014}
Milshteyn, A., Schneider, J., and Brady, S. (2014).
\newblock Mining the {Metabiome}: {Identifying} {Novel} {Natural} {Products}
  from {Microbial} {Communities}.
\newblock {\em Chemistry \& Biology}, 21(9):1211--1223.

\bibitem[Mueller and Sachs, 2015]{mueller_engineering_2015}
Mueller, U.~G. and Sachs, J.~L. (2015).
\newblock Engineering {Microbiomes} to {Improve} {Plant} and {Animal} {Health}.
\newblock {\em Trends in Microbiology}, 23(10):606--617.
\newblock Publisher: Elsevier.

\bibitem[Ofiţeru et~al., 2010]{ofiteru_combined_2010}
Ofiţeru, I.~D., Lunn, M., Curtis, T.~P., Wells, G.~F., Criddle, C.~S.,
  Francis, C.~A., and Sloan, W.~T. (2010).
\newblock Combined niche and neutral effects in a microbial wastewater
  treatment community.
\newblock {\em Proceedings of the National Academy of Sciences},
  107(35):15345--15350.
\newblock Publisher: Proceedings of the National Academy of Sciences.

\bibitem[Panke-Buisse et~al., 2015]{panke-buisse_selection_2015}
Panke-Buisse, K., Poole, A.~C., Goodrich, J.~K., Ley, R.~E., and Kao-Kniffin,
  J. (2015).
\newblock Selection on soil microbiomes reveals reproducible impacts on plant
  function.
\newblock {\em The ISME Journal}, 9(4):980--989.

\bibitem[Pflughoeft and Versalovic, 2012]{pflughoeft_human_2012}
Pflughoeft, K.~J. and Versalovic, J. (2012).
\newblock Human {Microbiome} in {Health} and {Disease}.
\newblock {\em Annual Review of Pathology: Mechanisms of Disease},
  7(1):99--122.
\newblock \_eprint: https://doi.org/10.1146/annurev-pathol-011811-132421.

\bibitem[Popovic, 2004]{popovic_asymptotic_2004}
Popovic, L. (2004).
\newblock Asymptotic genealogy of a critical branching process.
\newblock {\em The Annals of Applied Probability}, 14(4):2120--2148.

\bibitem[Postek et~al., 2022]{postek_microfluidics_2022}
Postek, W., Pacocha, N., and Garstecki, P. (2022).
\newblock Microfluidics for antibiotic susceptibility testing.
\newblock {\em Lab on a Chip}, 22(19):3637--3662.

\bibitem[Potvin-Trottier et~al., 2018]{potvin-trottier_microfluidics_2018}
Potvin-Trottier, L., Luro, S., and Paulsson, J. (2018).
\newblock Microfluidics and single-cell microscopy to study stochastic
  processes in bacteria.
\newblock {\em Current opinion in microbiology}, 43:186--192.

\bibitem[Sniegowski et~al., 2000]{sniegowski_evolution_2000}
Sniegowski, P.~D., Gerrish, P.~J., Johnson, T., and Shaver, A. (2000).
\newblock The evolution of mutation rates: separating causes from consequences.
\newblock {\em BioEssays: News and Reviews in Molecular, Cellular and
  Developmental Biology}, 22(12):1057--1066.

\bibitem[Sniegowski et~al., 1997]{sniegowski_evolution_1997}
Sniegowski, P.~D., Gerrish, P.~J., and Lenski, R.~E. (1997).
\newblock Evolution of high mutation rates in experimental populations of {E}.
  coli.
\newblock {\em Nature}, 387(6634):703--705.

\bibitem[Swenson et~al., 2000]{swenson_artificial_2000}
Swenson, W., Wilson, D.~S., and Elias, R. (2000).
\newblock Artificial ecosystem selection.
\newblock {\em Proceedings of the National Academy of Sciences},
  97(16):9110--9114.

\bibitem[Toprak et~al., 2013]{toprak_building_2013}
Toprak, E., Veres, A., Yildiz, S., Pedraza, J.~M., Chait, R., Paulsson, J., and
  Kishony, R. (2013).
\newblock Building a {Morbidostat}: {An} automated continuous-culture device
  for studying bacterial drug resistance under dynamically sustained drug
  inhibition.
\newblock {\em Nature protocols}, 8(3):555--567.

\bibitem[Trivedi et~al., 2016]{trivedi_response_2016}
Trivedi, P., Delgado-Baquerizo, M., Anderson, I.~C., and Singh, B.~K. (2016).
\newblock Response of {Soil} {Properties} and {Microbial} {Communities} to
  {Agriculture}: {Implications} for {Primary} {Productivity} and {Soil}
  {Health} {Indicators}.
\newblock {\em Frontiers in Plant Science}, 7.

\bibitem[Van~den Bergh et~al., 2018]{van_den_bergh_experimental_2018}
Van~den Bergh, B., Swings, T., Fauvart, M., and Michiels, J. (2018).
\newblock Experimental {Design}, {Population} {Dynamics}, and {Diversity} in
  {Microbial} {Experimental} {Evolution}.
\newblock {\em Microbiology and Molecular Biology Reviews},
  82(3):10.1128/mmbr.00008--18.

\bibitem[Wahl and Gerrish, 2001]{wahl_probability_2001}
Wahl, L.~M. and Gerrish, P.~J. (2001).
\newblock The {Probability} {That} {Beneficial} {Mutations} {Are} {Lost} in
  {Populations} with {Periodic} {Bottlenecks}.
\newblock {\em Evolution}, 55(12):2606--2610.

\bibitem[Wahl et~al., 2002]{wahl_evaluating_2002}
Wahl, L.~M., Gerrish, P.~J., and Saika-Voivod, I. (2002).
\newblock Evaluating the impact of population bottlenecks in experimental
  evolution.
\newblock {\em Genetics}, 162(2):961--971.

\bibitem[Wahl and Krakauer, 2000]{wahl_models_2000}
Wahl, L.~M. and Krakauer, D.~C. (2000).
\newblock Models of experimental evolution: the role of genetic chance and
  selective necessity.
\newblock {\em Genetics}, 156(3):1437--1448.

\bibitem[Wahl and Zhu, 2015]{wahl_survival_2015}
Wahl, L.~M. and Zhu, A.~D. (2015).
\newblock Survival {Probability} of {Beneficial} {Mutations} in {Bacterial}
  {Batch} {Culture}.
\newblock {\em Genetics}, 200(1):309--320.

\bibitem[Wilson, 1975]{wilson_theory_1975}
Wilson, D.~S. (1975).
\newblock A theory of group selection.
\newblock {\em Proceedings of the National Academy of Sciences},
  72(1):143--146.

\bibitem[Xie and Shou, 2021]{xie_steering_2021}
Xie, L. and Shou, W. (2021).
\newblock Steering ecological-evolutionary dynamics to improve artificial
  selection of microbial communities.
\newblock {\em Nature Communications}, 12(1):6799.

\bibitem[Xie et~al., 2019]{xie_simulations_2019}
Xie, L., Yuan, A.~E., and Shou, W. (2019).
\newblock Simulations reveal challenges to artificial community selection and
  possible strategies for success.
\newblock {\em PLOS Biology}, 17(6):e3000295.

\end{thebibliography}

\section*{Acknowledgements}

GD gratefully acknowledges the financial support of the \emph{Origines et Conditions d’Apparition de la Vie} (OCAV) programme, PSL University (ANR-10-IDEX-001–02) during the elaboration of the manuscript and of the John Templeton Foundation (\#62220) during its revision. 

AL thanks the \emph{Center for Interdisciplinary Research in Biology} (CIRB, Coll\`ege de France) for funding.

This work has been inspired by discussions with Paul B. Rainey and Silvia De Monte during the OCAV programme.

\section*{Reproducing the figures}

The code used to generate all the figures in the manuscript uses python 3.11, matpolotlib 3.6.3, numpy 1.24.2 and scipy 1.24.2. It is available on the Zenodo repository under the DOI \href{https://doi.org/10.5281/zenodo.8087961}{10.5281/zenodo.8087961}. 

\markright{Appendix}

In the following sections we give the mathematical proofs for the results used in the main text.

\section*{Parameters range}

\begin{proof}[Proposition \ref{lem:survival_prob} - Survival Probability]
\label{proof:survival_prob}
Let $s_n$ be the survival probability at $t=nT$ of a lineage started from a single cell at $t=0$. 

\textbf{\textbf{First cycle}}

Consider one cell at time $t=0$. This cell follows a Linear Markov Branching
Process $(Z_t)_{t\in [0,T)}$ with constant rates $b,d$ until the dilution time $T$. We use two results from \cite{athreya_branching_1972}. First, The branching process goes extinct ($Z_t=0$) with probability $q(b,d,t):=\mathbb P(Z_t = 0)$. Second, conditional on non-extinction $Z_t$ follows a geometric distribution with parameter
$p(b,d,t)$, that is, $\mathbb P(Z_t = k | Z_t \neq 0)=p(1-p)^{k-1}$. The values of $p$ and $q$ are given by the following table:

\begin{center}
\begin{tabular}{llll}
\hline
 &  & $p(b,d,t)$ & $q(b,d,t)$\\
\hline
Subcritical cells & $r < 0$ & $\frac{d-b}{d-be^{rt}}$ & $\frac{d(1-e^{rt})}{d-be^{rt}}$\\
Critical cells & $r = 0$ & $\frac{1}{1+bt}$ & $\frac{bt}{1+bt}$\\
Supercritical cells & $r > 0$ & $\frac{(b-d)e^{-rt}}{b-de^{-rt}}$ & $\frac{d(1-e^{-rt})}{b-de^{-rt}}$\\
\hline
\end{tabular}
\end{center}

The probability generating function of $Z_T$ is:
\begin{align*}
f_{Z_T}(s) &=  \sum_{k\geq 0}\mathbb P(Z_T = k) s^k &\text{Definition of } f\\
&= \mathbb P(Z_T = 0)s^0 + \sum_{k\geq 1}\mathbb P(Z_T = k) s^k\\
&= \mathbb P(Z_T = 0)s^0 + \sum_{k\geq 1}\mathbb P(Z_T \neq 0)P(Z_T = k | Z_T \neq 0) s^k\\
&= q + \sum_{k\geq 1} (1-q)p(1-p)^{k-1} s^k &\text{Definition of } p,q\\
&= q + ps(1-q)\sum_{k\geq 0} ((1-p)s)^k\\
%&= q + \frac{p(1-q)}{1-p} \sum_{k\geq 1} ((1-p)s)^k\\
%&= q + \frac{p(1-q)}{1-p} \left (- 1 + \sum_{k\geq 0} ((1-p)s)^k \right )\\
%&= q + \frac{p(1-q)}{1-p} \left (- 1 + \frac{1}{1-(1-p)s} \right )
&= q + \frac{ps(1-q)}{1-(1-p)s}
&\text{Geometric series, } (1-p)s<1\\
%&= q + \frac{p(1-q)}{1-p} \frac{s(1-p)}{s(p-1)+1}\\
%&= q + \frac{sp(1-q)}{s(p-1)+1}\\
&= \frac{s(p-q)+q}{s(p-1)+1}\\
\end{align*}

As expected, $f_{Z_T} (1) = 1$  and $f_{Z_T} (0) = P(Z_T = 0) = q$.\\

\textbf{\textbf{\textbf{Second cycle}}}

Consider a cell at the end of the first cycle, just before the first dilution. Let $B$ be the number of descendants of this cell at the end of the second cycle. There are two possibilities to consider: 

\begin{itemize}
    \item with probability $(1-\delta)$, the cell is discarded during the dilution at the beginning of the second cycle and $B=0$,
    \item with probability $\delta$, the cell is not discarded, and thus sees its descent grow until the end of the cycle following the same law as in the first circle, namely the law of $Z_T$. 
\end{itemize}

Thus,\begin{equation*}f_B = 1-\delta +\delta f_{Z_T}.\end{equation*}

We now introduce a notation for the rest of the proof: $f_{Z_T}$ and $f_B$ are linear fractional, a property that will be useful later when composing generating functions. We associate any linear-fractional function with a coefficient matrix $M$ as follows:
\begin{equation*}
\forall M = \begin{bmatrix} v & w \\ x & y \end{bmatrix} \in M_2(\mathbb R),\quad h(M,\cdot) : s\mapsto h(M,s) = \frac{vs+w}{xs +y}.
\end{equation*}

We define $Q_\varepsilon$ for all $\varepsilon$ in [0,1) as:
\begin{equation*}
Q_\varepsilon := \begin{bmatrix}
p-1 +\varepsilon(1-q)& 1-\varepsilon(1-q) \\
p-1 & 1
 \end{bmatrix}.\end{equation*}
Note that, $f_{Z_T}$ is associated with the matrix $Q_1$ and $f_B$ with the matrix $Q_\delta$.

\textbf{\textbf{\textbf{Subsequent cycles}}}

Let $X_n$ be the number of descendants of an ancestral cell at the end of the $n$th cycle. Let $(B_{n,i})_{n>0, i>0}$ be a collection of identical and independent random variables following the same law as $B$. Then the following recursion holds:
\begin{equation*}
    X_n = \sum_{i=1}^{X_{n-1}} B_{n,i} 
\end{equation*}

Since $B$ and $X_{n-1}$ are independent, the generating function of $X_n$ is given by the composition of the generating functions of $X_{n-1}$ and $B$: $f_{X_n} = f_{X_{n-1}} \circ f_B $. 

By induction, the generating function of $X_n$ is obtained by composing the generating function of $x_1$ with the $n-1$ times iteration of the generating function of $B$: 
\begin{equation*}f_{X_n} = f_{X_1} \circ \underbrace{f_B \circ \ldots \circ f_B}_{(n-1)\text{ times}}.\end{equation*}

Now, recall that $X_1$ follows the same law as $Z_t$ (associated with the matrix $Q_1$) and $f_B$ is associated with the matrix $Q_\delta$. From this observation and the composition rule outlined in the previous section, it results that: 
\begin{equation*}
f_{X_n} = h(Q_1Q_\delta^{n-1}, \cdot)
\end{equation*}

It can be checked that this expression corresponds to $n$ cycles, with $n-1$ dilutions by a factor $\delta$, since there is no dilution before the first cycle.

This leads to the survival probability of the lineage spawned by an ancestral cell, at the end of cycle $n$:

\begin{equation*}
s_n = 1 - f_{X_n}(0) = 1 - h(Q_1Q_\delta^{n-1},0)
\end{equation*}
\end{proof}

\begin{proof}[Proposition \ref{lem:survival_prob_long_term} - Long Term Survival Probability]
\label{proof:survival_prob_long_term}
Let us look for the fixed points of $f_{B}$:
\begin{align*}
f_{B}(s) = s &\Leftrightarrow (p-1+\delta(1-q))s + 1-\delta(1-q) = s ((p-1)s+1)\\
&\Leftrightarrow (s-1)((p-1)s-\delta(1-q)-1) = 0\\ 
&\Leftrightarrow s = \frac{\delta(1-q)-1}{p-1} &\text{or } s=1
\end{align*}
$f_{B}(s)-s$ is a second order polynomial for $s$ meaning that $f_{B}$ has at most two fixed points. 
One of them is 1, in accordance with the definition of characteristic functions. 
The other is $s^*$, with:
\begin{equation*}
s^* < 1 \Leftrightarrow p   < \delta(1-q).
\end{equation*}

Since $f_B$ is continuous, $f_B(s)\ge s$ for all $s\in [0, \min (s^*, 1)]$ and $\min (s^*, 1)$ is a fixed point of $f_B$, the
sequence $u_n = f_{B}(u_{n-1})$ with $u_0 = h(Q_1,0) \in [0,1]$
converges to $\min (s^*, 1)$, that is, to $1$ if $p > \delta(1-q)$ and to $s^*$ otherwise.

\begin{itemize}
\item In \textbf{Subcritical regime}, $p   < \delta(1-q) \Leftrightarrow e^{-rT}<\delta$, is always false since $e^{-rT}>1$ and $\delta \leq 1$.
\item In \textbf{Critical regime}, $p < \delta(1-q) \Leftrightarrow \frac{1}{1+bT} < \delta \left(1-\frac{bT}{1+bT} \right) = \delta \frac{1}{1+bT}$, is always false since $\delta \leq 1$.
\item In \textbf{Supercritical regime}, $p   < \delta(1-q) \Leftrightarrow e^{-rT}<\delta$,
\end{itemize}

Thus, $s^*>1$ except if $b>d$ and $\delta > e^{-rT}$. In this case,

\begin{align*}
s_\infty &= 1 - h(Q_1,s^*) \\
&= \frac{p-\delta(1-q)}{\delta(p-1)}
\\&= \frac{r(\delta - e^{-rT})}{\delta b (1-e^{-rT})}
%1 - \frac{\delta(1-q)-1}{p-1}  \\
%&= \frac{\left(\delta - e^{T \left(- b + d\right)}\right) \left(b - d\right)}{b \left(1 - e^{T \left(- b + d\right)}\right)}\\
%&= \frac{r(\delta-e^{-rT})}{b (1-e^{-rT})} 
\end{align*}
\end{proof}

\begin{proof}[Proposition \ref{lem:snstar} - Critical survival Probability]
\label{proof:snstar}

Let  $r=b-d$, $r>0$, we use the expression of $p$ and $q$ from the proof of Proposition \ref{lem:survival_prob} and let $\delta=e^{-rT}$.  Then, the probability of extinction of the branching process is:
\begin{equation*}
q = \frac{d(1-\delta)}{b-d\delta}
\end{equation*}

The parameter of the geometric distribution of the number of cells, conditional on non extinction, is:
\begin{equation*}
p = \frac{r\delta}{b-d\delta}
\end{equation*}
Thus, it is possible to rewrite $Q_1$ and $Q_\delta$ like this: 

\begin{align*}
Q_1 = \frac{1}{b-\delta d} \begin{bmatrix}
\delta b-d & d(1-\delta) \\
b(\delta-1) & b-\delta d
 \end{bmatrix} := \frac{1}{b-\delta d} \overline{Q_1} \\
 Q_\delta = \frac{1}{b-\delta d} \begin{bmatrix}
 2\delta b -\delta d -b & b(1-\delta) \\
b(\delta-1) & b-\delta d
 \end{bmatrix} := \frac{1}{b-\delta d} \overline{Q_\delta} \\
\end{align*}

By induction, we can show that:
\begin{align*}
s_n^* &= 1-h(Q_1Q_\delta^{n-1},0) = 1-\frac{(Q_1Q_\delta^{n-1})_{01}}{(Q_1Q_\delta^{n-1})_{11}} 
=  1-\frac{\overline{Q_1Q_\delta^{n-1}}_{01}}{\overline{Q_1Q_\delta^{n-1}}_{11}}\\
&= \frac{b-d}{bn-\delta(b(n-1)+d)} = \frac {r}{bn(1-\delta)+\delta r} 
\end{align*}

\end{proof}
\section*{Coalescent Point Processes}

\begin{proof}[Proposition \ref{lem:coal_lineage} - Coalescent tree]
\label{proof:coal_lineage}

First, we recall the Proposition 7 from \cite{lambert_birthdeath_2013}: \begin{quote}
 ``Start with a CPP tree with inverse tail distribution $F$. Add extra mass extinctions with survival probabilities $\epsilon_1,\ldots\epsilon_k$ at times $T-s_1 > \ldots > T-s_k$ (where $s_1>0$ and $s_k<T$). Then, conditional on survival, the reconstructed tree of the phylogenetic tree obtained after the passage of mass extinctions is again a coalescent point process with inverse tail distribution $F_\epsilon$ given by \begin{equation*}
     F_\epsilon(t) = \epsilon_1 \ldots \epsilon_m F(t) + \sum_{j=1}^m(1-\epsilon_j)\epsilon_1 \ldots \epsilon_{j-1}F(s_j),\quad t \in [s_m,s_{m+1}], m\in \{0,1,\ldots,k\},
 \end{equation*} where $s_0:=0$ and $s_{k+1}=T$ (empty sum is zero, empty product is 1).''
 \end{quote}

Consider the CPP of the population just before the $n$th cycle of dilution. The population experienced $n-1$ bottlenecks at times $t_1=(n-1)T, \ldots, t_{n-2} = 2T, t_{n-1} = T$.

Let $\tilde F$ be the scale function of the CPP without bottlenecks. Using Proposition 7 in \cite{lambert_birthdeath_2013} with $\epsilon_i = \delta$ and $s_j=jT$, it is possible to write $F$ as: \begin{equation*}
     F(t) = \delta^k \tilde F(t) + (1-\delta)\sum_{j=1}^k\delta^{j-1}\tilde F(jT),\quad t \in [kT,(k+1)T], k\in \{0,1,\ldots,n\},
 \end{equation*} 
%\begin{equation*}
%F(t) = \delta^{n} \tilde F (nT+s) + (1-\delta) \sum_{j=0}^{n-1}  \delta^{j} \tilde F (jT),\; t \in %[kT,(k+1)T], k\in %\{0,1,\ldots,n\},
%\end{equation*}
%\begin{equation*}
%F(t) = \delta^{n} \tilde F (nT+s) + (1-\delta) \sum_{j=0}^{n-1}  \delta^{j} \tilde F (jT)
%\end{equation*}
If the sampling at the last cycle is taken into account, the scale function becomes $\overline{F}$:

\begin{equation*}
\overline{F}(t) =  1 - \delta + \delta F(t)
\end{equation*}
The expression of $\tilde F$ is known for the most common birth-death
processes:

\begin{center}
\begin{tabular}{lll}
 & \textbf{Parameters} & \textbf{Scale function}\\
\hline
 &  & \\
Pure Birth & $0 = d < b$ & $\tilde F(t) = e^{bt}$\\
Non-critical & $0 < d \neq b > 0$ & $\tilde F(t) = \frac{b}{r}(e^{rt} - 1) + 1$\\
Critical & $0 < d = b$ & $\tilde F(t) = bt+1$\\
 &  & \\
\end{tabular}
\end{center}

\noindent\textbf{\textbf{Non-critical case}} Let $r := b-d \neq 0, b>0, d\geq0$, $k \in \{0, \ldots n\}$, $s<T$:
\footnotesize
\begin{align*}
F(kT+s) &=
\delta^k \left [1 - \frac br + e^{r(kT+s)} \frac br \right]
+ (1-\delta) \sum_{j=1}^{k} \delta^{j-1} \left [ e^{rjT} \frac br + 1 - \frac br \right]\\
&= \delta^k \left (1 - \frac br \right ) + \frac br e^{rs} (\delta e^{rT})^k
+ (1-\delta)
\left [\left ( 1-\frac br \right ) \sum_{j=0}^{k-1} \delta^j
+ \frac br e^{rT} \sum_{j=0}^{k-1} (\delta e^{rT})^j \right ]\\
&= \delta^k \left (1 - \frac br \right ) + \frac br e^{rs} (\delta e^{rT})^k
+ \left ( 1-\frac br \right ) (1-\delta^k)
+ \frac {b (1-\delta)}{r} e^{rT} \sum_{j=0}^{k-1} (\delta e^{rT})^j
& \text{(Geometric series, } \delta \neq 1)\\
&= 1 + \frac br \left[ e^{rs}(\delta e^{rT})^k -1
+ (1-\delta) e^{rT} \sum_{j=0}^{k-1} (\delta e^{rT})^j \right ]\\
\end{align*}
\normalsize

If $\delta \neq \delta^*$ then,
\begin{align*}
F(kT+s) &=
1 + \frac br \left[ e^{rs}(\delta e^{rT})^k -1
+  (1-\delta) e^{rT} \frac{1-(\delta e^{rT})^k}{1-\delta e^{rT}}   \right ] & \text{(Geometric series, } \delta^*e^{rT} \neq 1)\\
\end{align*}

Otherwise, if $\delta = \delta^* = e^{-rT}$,
\begin{align*}
F(kT+s) =
1 + \frac br \left (e^{rs} - 1 +  k(e^{rT}-1) \right )
\end{align*}

\noindent\textbf{\textbf{Critical case}} $b=d\neq 0$
\begin{align*}
F(kT+s) =& \delta^k (b(kT + s) + 1) + (1-\delta) \sum_{j=1}^{k} \delta^{j-1} (bjT + 1)\\
=&\delta^k (b(kT + s) + 1) + (1-\delta) \sum_{j=0}^{k-1} \delta^{j} (b(j+1)T + 1)\\
=&\delta^k (b(kT + s) + 1) + (1-\delta) \left [ (1+bT) \sum_{j=0}^{k-1} \delta^{j} + b T \sum_{j=0}^{k-1} j\delta^{j} \right ]\\
=&\delta^k (b(kT + s) + 1) + (1-\delta^k) (1+bT) + (1-\delta) b T \sum_{j=0}^{k-1} j\delta^{j}
& \text{(Geometric series, } \delta \neq 1)\\
=&\delta^k b(T(k - 1)+s) + 1+bT + (1-\delta) b T \sum_{j=0}^{k-1} j\delta^{j}\\
=&1+\delta^k b(T(k - 1)+s) +  b T  \left [1 + (1-\delta) \sum_{j=0}^{k-1} j\delta^{j} \right ]\\
=&1 + b\frac{T-s\delta^{k+1}-(T-s)\delta^k}{1-\delta} & (\delta \neq 1)\\
\end{align*}
\end{proof}

\begin{proof}[Proposition \ref{lem:length_coal} - Length of the coalescent tree]
\label{proof:length_coal}
Let $M_n$ be the expected number of mutations within the serial transfer experiment at the end of the $n$th cycle. Let also $T$ be the duration of the growth phase and $\delta=\delta^*$ be the dilution factor. Suppose that the first cycle is seeded with a single ancestral cell. The initial lineage survives to cycle $n$ with probability $s_n^*$ (since $\delta=\delta^*$, see Proposition \ref{lem:snstar}). 

Each extant lineage after $n$ cycles spawns an independent coalescent tree with expected length $L_n$. Moreover, mutations are accumulated following a Poisson point process on the tree with intensity $\theta$. Hence, the expected number of mutations accumulated on one tree is $\theta L_n$. 

Thus, 
\begin{equation*}
    M_n = \theta s_n^* L_n.
\end{equation*}

\bigskip 

Let us now compute $L_n$. 

Let $\{H_{ij}, (i,j) \in \mathbb N^2\}$ be a set of i.i.d. random
variables following the same law as $H$, defined by its inverse tail
distribution $F(t)=\frac{1}{\mathbb P(H > t)}$.

Let $\tau_n$ be a CPP with branches $H$ stopped at $nT$.
Let $N(\tau_n)$ be the number of leaves of the random tree $\tau_n$.
The length of the tree  $\mathcal L(\tau_n)$ is the random variable:
\begin{equation*}
\mathcal L(\tau_n) = nT + \sum_{j=1}^{N(\tau_n)-1} H_j,
\end{equation*}
where $nT$ is the length of the spine and $H_j$ are the length of the other $N(\tau_n)-1$ branches. 

Its expected value is:
\begin{equation*}
L_n = \mathbb E (\mathcal L(\tau_n)) = nT + (\mathbb E(N(\tau_n))-1) \mathbb E(H|H < nT)
\end{equation*}

\textbf{\textbf{Number of leaves}}: $N(\tau_n)$ is a geometric random variable:

\begin{equation}
\label{eq:probanumberofleaves}
\mathbb P(N(\tau_n ) = k) = P(H>nT)P(H\leq nT)^{k-1} = \frac{1}{F(nT)} \left(1 - \frac{1}{F(nT)} \right )^{k-1}
\end{equation}

Indeed, if $k$ is the index of the first $H_i$ such that $H_i>nT$, there are $k-1$ branches, plus the spine, for a total of $k$ leaves. 

Thus, the expected number of leaves of $\tau_n$, is:
\begin{equation}
\label{eq:expected_leaves_cpp}
\mathbb E (N( \tau_n )) = F(nT).
\end{equation}

Note that 
\begin{equation*}
\mathbb E (N( \tau_n) -1) = F(nT) - 1 = \frac{1-\frac{1}{F(nT)}}{\frac{1}{F(nT)}} = \frac{\mathbb P(H<nT)}{\mathbb P(H>nT)},
\end{equation*}
so we get
\begin{equation*}
L_n = nT + \frac{\mathbb P(H<nT)}{\mathbb P(H>nT)}\mathbb E(H| H < nT) %= nT+ F(nT) \mathbb E(H, H < nT)
\end{equation*}

\textbf{\textbf{Length of branches}}:

We recall that for a positive r.v. $X$, $\mathbb E(X) = \int_0^{+\infty}  \mathbb P(X>x) dx$.

We can rescale the tail-distribution to take into account the conditioning:

\begin{equation}
\label{eq:conditional_proba}
\mathbb P(H>x | H<y) = \begin{cases}
 0  & \text{ if } x>y,\\
 \frac{\mathbb P(H>x) - \mathbb P(H>y)}{\mathbb P(H<y)} & \text{ otherwise.}
\end{cases}
\end{equation}

Thus,
\begin{align*}
\mathbb E (H | H<nT) &=   \int_0^{nT} \mathbb P(H>x | H<nT) dx \\
&= \int_0^{nT}  \frac{\mathbb P(H>x) - \mathbb P(H>nT)}{\mathbb P(H<nT)}dx \\
&= \frac{1}{\mathbb P(H<nT)} \left [ \int_0^{nT} \frac{dx}{F(x)} - \frac{nT}{F(nT)} \right ] & \left ( \text{By definition, } \mathbb P(H>x) =: \frac{1}{F(x)} \right ),
%&= \frac{F(nT)}{F(nT) - 1} \left [\sum_{k=0}^n \int_{kT}^{(k+1)T} \frac{dx}{F(x)} - \frac{nT}{F(nT)} \right ]
\end{align*}
so that
\begin{align}
L_n %&=  nT + \mathbb E(N(\tau_n)-1) \mathbb E(H|H < nT)\nonumber \\
&= nT + \frac{\mathbb P(H<nT)}{\mathbb P(H>nT)} \mathbb E(H|H < nT)\nonumber\\
&= nT + F(nT) \left [ \int_0^{nT} \frac{dx}{F(x)}   - \frac{nT}{F(nT)}\right] \nonumber \\
&=  F(nT) \int_0^{nT} \frac{dx}{F(x)}  \label{eq:ln_f}
\end{align}

Moreover, $\forall k \in \mathbb N$ and $\forall s \in \mathbb R,  s<T$:
\begin{equation*}
F(kT+s) = 1 + \frac br (k(e^{rT}-1) + e^{rs} - 1),
\end{equation*}
so that
\begin{align*}
\int_{kT}^{(k+1)T} \frac{dx}{F(x)}  &= \int_{0}^{T} \frac{1}{1 + \frac br (k(e^{rT}-1) + e^{rs} - 1)} ds \\
&= \frac{rT - \log \left ( \frac{k (e^{rT} - 1) + \frac rb }{(k+1) (e^{rT} - 1) + \frac rb } \right ) }{ bk(e^{rT}-1)-d}
\end{align*}

Finally, we get:
\begin{equation}
    L_n = \left ( 1 + \frac br n (e^{rT} - 1) \right ) \sum_{k=0}^n  \frac{rT - \log \left ( \frac{k (e^{rT} - 1) + \frac rb }{(k+1) (e^{rT} - 1) + \frac rb } \right ) }{ bk(e^{rT}-1)-d} 
\end{equation}
\end{proof}

\begin{proof}[Proposition \ref{lem:mutatation_frequency_spectrum} - Mutation Frequency Spectrum]
\label{proof:mutation_frequency_spectrum}
We now turn to a more detailed account of the mutation distribution.

\textbf{Fixed and segregating mutations}

Mutations compared to the ancestral type are either fixed (shared by all individuals) or segregating (not shared by all individuals). Thus, we have:
\begin{equation*}
    M^f_n + M^s_n = M_n
\end{equation*}

Fixed mutations are the ones found between the root of the coalescent tree ($t=nT$) and the first branching. Let $Y$ be the random variable encoding this length. We have $Y=nT-X$ with $X$ the random variable encoding the height of the first branching. Thus,
\begin{equation*}
    M^f_n = \theta s_n \mathbb E(Y) = \theta s_n (nT-\mathbb E(X)),
\end{equation*}
where $\theta$ is the mutation rate, $s_n$ is the probability that the lineage spawned by one of the ancestral cells is still extant at the end of the $n$th cycle. 

Now we compute $\mathbb E(X)$. Let $H_1...H_{N(\tau_{n})-1}$ be the branch lengths of the CPP stopped at $nT$. The variables $(H_i)$ are independent and identically distributed as $H$, conditioned to be smaller than $nT$. X is the maximum of those values:
\begin{equation*}
    X := \max\left\{H_1, H_2... H_{N(\tau_{n})-1}\right\}.
\end{equation*}

Note that the first branching cannot be higher than the length of the spine. Thus, $\mathbb P(X>nT)=0$. Otherwise:

\begin{align*}
\mathbb P(X>s) &= 1 - P(X<s) \\
&= 1 - \sum_{k=1}^{\infty} \mathbb P(N(\tau_n)=k) \mathbb{P}(X<s|N(\tau_n)=k) &\text{(Total probability)}\\
&= 1 -\sum_{k=1}^{\infty} \mathbb P(N(\tau_n)=k) \mathbb P (H<s | H<nT)^{k-1} & (X \text{is the max of } (H_i) \text{ i.i.d.)}\\
     &= 1- \sum_{k=1}^{\infty} \frac{1}{F(nT)}\left(1-\frac{1}{F(nT)} \right)^{k-1} \left( 1 -  \frac{\frac{1}{F(s)}-\frac{1}{F(nT)}}{1-\frac{1}{F(nT)}} \right)^{k-1} & \text{(Eq. \ref{eq:probanumberofleaves} and \ref{eq:conditional_proba})}.\\
    &= 1- \sum_{k=1}^{\infty} \frac{1}{F(nT)}\left(1-\frac{1}{F(nT)} \right)^{k-1} \left(   \frac{1-\frac{1}{F(s)}}{1-\frac{1}{F(nT)}} \right)^{k-1}\\
     &= 1- \frac{1}{F(nT)}
     \sum_{k=0}^\infty \left (1 - \frac{1}{F(s)} \right )^{k} \\
     &= 1- \frac{1}{F(nT)} \frac{1}{1-1+\frac{1}{F(s)}} &\text{(Geometric series)}\\
     &= 1 - \frac{F(s)}{F(nT)}
\end{align*}

Again, we recall that for a positive r.v. $X$, $\mathbb E(X) = \int_0^{+\infty}  \mathbb P(X>x) dx$. Thus, 
\begin{equation*}
    \mathbb E(X) = \int_0^{nT} 1 - \frac{F(s)}{F(nT)} ds = nT - \int_0^{nT} \frac{F(s)}{F(nT)} ds.
\end{equation*}
Thus, the expected number of fixed mutations is:
\begin{equation*}
    M_n^f = \theta s_n [nT - \mathbb E(X)] = \theta s_n \int_0^{nT}  \frac{F(s)}{F(nT)} ds
\end{equation*}
Finally, the expected number of segregating mutations within the droplet is the total number of mutations ($M_n$) minus the number of fixed mutations: 
\begin{align*}
 M_n^s &= M_n - M_n^f \\
 &= \theta s_n \int_0^{nT} \frac{F(nT)}{F(s)} - \frac{F(s)}{F(nT)} ds 
\end{align*}

\bigskip 
\textbf{Full spectrum}
For a sample of $v$ individuals in a CPP stopped at time $nT$, Theorem 2.2 in \cite{lambert_allelic_2009} gives the expression of $\mathbb E(S_v(u))$ the expected number of mutant sites that are carried by exactly $1\leq u \leq v-1$ individuals.
\begin{equation}
    \mathbb E (S_v(u)) = \theta \int_0^{nT} \left(1-\frac{1}{W(x)} \right)^{u-1} \left ( \frac{v-u-1}{W(x)^2} + \frac{2}{W(x)} \right) dx,
\end{equation}
with $W$ the inverse tail distribution of the branch length in the stopped CPP. 
$W$ is expressed, for $x<nT$, as a rescaling of $F$: 
\begin{equation}
    W(x) = \frac{1}{\mathbb P(H>x | H<nT)} = \frac{1 - \frac{1}{F(nT)}}{\frac{1}{F(x)}-\frac{1}{F(nT)}}.
\end{equation}

The expected value of $S_v(u)$ converges towards a limit \citep{lambert_allelic_2009} when the sample size increases, provided $\mathbb E(H|H<nT)<\infty$, which is the case here, since the CPP is stopped at $nT$.

\begin{equation}
    \lim_{v\to\infty} v^{-1} \mathbb E (S_v(u)) = \theta \int_{0}^{nT} \left(1-\frac{1}{W(x)} \right)^{u-1} 
    \frac{1}{W(x)^2} dx 
\end{equation}
Thus, 
\begin{equation}
   a_u =  \lim_{v\to\infty} v^{-1} \mathbb E (S_v(u)) = \theta \int_0^{nT} \left(1- \frac{\frac{1}{F(x)}-\frac{1}{F(nT)}}{1 - \frac{1}{F(nT)}} \right)^{u-1} 
    \left ( \frac{\frac{1}{F(x)}-\frac{1}{F(nT)}}{1 - \frac{1}{F(nT)}}\right)^2 dx 
\end{equation}
\end{proof}

\section*{Split droplets}

\begin{proof}[Proposition \ref{lem:survival2drops} - Survival Probability (2 drops)]
\label{proof:survival2drops}

Let cells within droplets in serial transfer be modelled by a linear birth-death process with constant parameters $b$ and $d$, subject to periodic bottlenecks $\delta$ every duration $T$ for $m+n$ cycles. 
However, consider that at cycle $m$, $k$ droplets are produced (instead of one), and then each new droplet submitted independently to the serial transfer procedure for the remaining $n$ cycles. We call this situation $(k,m,n)$-split droplets. 

We recall from the proof of Proposition \ref{lem:survival_prob} (p.~\pageref{proof:survival_prob}) that the generating function of the Linear Markov Branching Process at time $T$ that is started with one individual with probability $\varepsilon$ or with zero particles with probability $1-\varepsilon$ is the linear fractional function $s \mapsto h({Q_\varepsilon},s)$ with the expression of $Q_\varepsilon$ given by Equation \ref{eq:Qdelta}.

Let $s_{k,m,n}$, the probability that a lineage spawned by a single cell is not extinct at the end of the $(n+m)$-th cycle of $(k,m,n)$-split droplets. 
In this situation, the cells survive surely before the first cycle ($\varepsilon=1$), then survive the dilution with probability $\delta$ before the cycles $2,3,\ldots(m-1)$, then survive with probability $k\delta$ before cycle $m$, and finally survive with probability $\delta$ again before cycles $(m+1), (m+2), \ldots, (m+n)$.

By using the composition rule established in the proof of Proposition  \ref{lem:survival_prob}, the survival probability can be computed using the following linear fractional function: \begin{equation*}
s_{k,m,n} = 1 - h(Q_1 Q^{m-1}_\delta Q_{k\delta} Q^{n-1}_\delta, 0)
\end{equation*}
It can be checked that this expression corresponds to $m+n$ cycles, one dilution with a factor $k\delta$ and $(m+n-2)$ dilutions with a factor $\delta$. 
\end{proof}

\begin{proof}[Proposition \ref{lem:total_diversity2drops} - Total Diversity (2 drops)]
\label{proof:total_diversity2drops}

Let $M_{k,m,n}$ be the expected number of unique mutations accumulated in a lineage at cycle $n+m$  after the splitting of the initial droplet at cycle $m$ into $k=1,2\ldots \left \lfloor \frac{1}{\delta}\right \rfloor$ droplets. Then:
\begin{equation*}
M_{k, m, n} = \theta s_m L_{k,m,n},
\end{equation*}
with $L_{k,m,n}$ the expected length of the coalescent tree of the population (conditional on survival) whose expression we will establish. 

Let $\mathcal L_{k,m,n}$ be the random variable encoding the length of the coalescent tree spawned by a single cell at cycle $m+n$, submitted to a bottleneck $\delta$ every $T$ unit of time, where at cycle $m$, the population was diluted into $k$ droplets instead of one, conditional on non extinction.

The length of this tree, conditioned on non extinction, is the sum of the length of the stump tree $\mathcal L^\dagger$ (i.e., the tree before the $m$th cycle) and the length of the corolla $\mathcal L^c$ (i.e., all the trees spawned by the lineages that are extant after the dilution at cycle $m$).\\

First, the stump tree is a CPP stopped at $Tm$ and sampled accordingly. Indeed, all the branches extant in $Tm$ are not necessarily still extant in the present at time $T(m+n)$. 

Let us define $\pi_{k,n}$ the probability that a branch extant at time $Tm$ will still be extant at time $T(m+n)$. It's the product of the probability that the lineage be sampled in the split ($k\delta$), and that it survives after the split ($s_n$): 
\begin{equation}
\pi_{k,n} = k \delta s_{n}
\end{equation}

We recall the Proposition 2 from \cite{lambert_birthdeath_2013}: \begin{quote}
``The genealogy of a Bernoulli(p)-sample taken from a CPP with inverse tail distribution $F$ is a CPP with typical node depth denoted $H_p$ with inverse tail distribution $F_p$ given by $F_p(t)=1-p+pF(t)$''. \end{quote}

The stump tree is a CPP with inverse tail distribution $F$, stopped at time $Tm$ and submitted to a Bernoulli sampling $\pi_{k,n}$. Thus, it is a CPP with inverse tail distribution: 
\begin{equation*}
F^\dagger := 1 - \pi_{k,n} + \pi_{k,n} F,
\end{equation*}
according to Proposition 2 from \cite{lambert_birthdeath_2013}. Its expected length,  $L^\dagger$, is (according to Equation \ref{eq:ln_f}):
\begin{equation*}
    \mathbb E(\mathcal L^\dagger) = L^\dagger =  \int_0^{mT} \frac{F^\dagger(mT)}{F^\dagger(s)}ds
\end{equation*}
Moreover, the expected number of leaves of the stump tree is (according to Equation \ref{eq:expected_leaves_cpp}): 
\begin{equation*}
 \mathbb E( N( \mathcal  L^\dagger)) = F^\dagger(mT).
\end{equation*}
Additionally, each leaf of the stump tree at time $Tm$ gives rise to a CPP of expected length $L_{n}$ (according to Equation \ref{eq:ln_f}):
\begin{equation*}
    \mathbb E(\mathcal L_{n}) = L_{n} =  \int_0^{nT} \frac{F(nT)}{F(s)}ds,
\end{equation*}
Thus, the expected length of the corolla is:\begin{equation}
\label{eq:length_corrola}
    \mathbb E(\mathcal L^c) = \mathbb E( N( \mathcal  L^\dagger)) \mathbb E(\mathcal L_{n}) = F^\dagger(mT) L_{n}.
\end{equation}

Finally, the expected length of the full tree is:
\begin{align*}
    L_{k,m,n}  
    =& \mathbb E (\mathcal L_{k,m,n}) \\
    =& \mathbb E(\mathcal L^\dagger) + \mathbb E(\mathcal L^c)\\
    =& L^\dagger + F^\dagger(mT) L_{n}\\
    =& \int_0^{mT} \frac{F^\dagger(mT)}{F^\dagger(s)}ds + F^\dagger(mT) \int_0^{nT} \frac{F(nT)}{F(s)}ds\\
    =&  F^\dagger(mT) \left [\int_0^{mT} \frac{1}{F^\dagger(s)}ds + \int_0^{nT} \frac{F(nT)}{F(s)}ds \right ]
\end{align*}
This leads to the final expression: 
\begin{equation*}
    M_{k, m, n} = \theta s_m  F^\dagger(mT) \left [\int_0^{mT} \frac{1}{F^\dagger(s)}ds + \int_0^{nT} \frac{F(nT)}{F(s)}ds \right ]
\end{equation*}
\end{proof}

\begin{proof}[Proposition \ref{lem:private_mutations} - Private Mutations - Split Droplet]
\label{proof:private_mutations}

Let a single droplet be split into $k=1\ldots\left \lfloor \frac{1}{\delta} \right \rfloor$ at cycle $m$. 
Let $M'_{k,m,n}$ be the expected number of mutations that are private to any of the $k$ droplets when observed at cycle $n+m$. To correctly enumerate all these mutations, we must add (as illustrated in Figure \ref{fig:nested_coal}) the ones that happen in the stump tree (before $m$) and in the corolla (after $m$):
\begin{equation*}
    M'_{k,m,n} = 
      \underbrace{ \mathcal S_{k,m,n}  }_{\text{stump}}+ 
      \underbrace{ \mathcal C_{k,m,n} }_{\text{corolla}}  
\end{equation*}

\textbf{Corolla}

All the mutations that happen in the corolla are private to a single droplet. 

The expected number of mutations that happen in the corolla is proportional to the expected length of the corolla forest $\mathcal L^c$ (given by Equation \ref{eq:length_corrola}), multiplied by the mutation rate. Thus,
\begin{align*}
    \mathcal C_{k,m,n} = \theta s_m \mathbb E (\mathcal L^c) = \theta s_m F^\dagger(mT) L_{n}\\
\end{align*}

\textbf{Stump Tree}

Mutations that happen in the stump tree can be carried by cell lineages present in one or several droplets (as explained in the Figure \ref{fig:nested_coal}). 

Let $\mathcal S_{k,m,n}$ be the expected number of mutations occurring in the stump tree (i.e.,~before cycle $m$) that are only carried by cells that are found in a single droplet. This number is proportional to the mutation rate $\theta$. Let us compute this value. 

The stump tree is a Coalescent Point Process stopped at $mT$, with inverse tail distribution $F$, sampled with probability $\pi_{k,n}=k\delta s_n$ at time $mT$. Thus, it has an inverse tail distribution $F^\dagger = 1-\pi_{k,n}+\pi_{k,n}F$. If the stump is not extinct (i.e.,~with probability $s_m$) it has a number of leaves $N(\mathcal  L^\dagger)$ that follows a geometric distribution with parameter $\frac{1}{F^\dagger(mT)}$.

Consider a mutation that occurred in the stump tree, and let $C$ be the random variable encoding the number of leaves of the stump tree that bear this mutation. This variable follows a probability distribution given in Theorem 2.2 of \cite{lambert_allelic_2009}.

Finally, if $C=j$, the probability that all the $j$ individuals carrying the focal mutation at the time of the split are sampled in the same droplet is $\frac{1}{k^{j-1}}$. 

Hence, by the formula of total probabilities : 
\begin{equation}
    \mathcal S_{k,m,n} = \theta s_m \sum_{i \geq 1} \mathbb P(N(\mathcal  L^\dagger))=i) \sum_{j=1}^i \mathbb P(C = j) \frac{1}{k^{j-1}}
\end{equation}

Let 
$$
W^\dagger(s) := [1-1/F^\dagger(mT)]/[1/F^\dagger(s)-1/F^\dagger(mT)]
$$
and for $\ell\ge 0$, let
$$
G_{\ell}(x) :=  \frac{\ell-1}{W^\dagger(x)^2} + \frac{2}{W^\dagger(x)} 
$$
if $\ell >0$, whereas $G_{0}(x) :=1$.
Then,
\begin{align*}
\mathcal S_{k,m,n} =&  
 \theta  s_{m}
\sum_{i\ge 1} \frac{1}{F^\dagger(mT)} \left (1 -  \frac{1}{F^\dagger(mT)}\right)^{i-1} 
\sum_{j=1}^{i} 
\int_0^{mT} dx \left ( 1 - \frac{1}{W^\dagger(x)} \right)^{j-1} G_{i-j}(x)\frac{1}{k^{j-1}} \\
=&\frac{\theta s_{m}}{F^\dagger(mT)}\int_0^{mT} dx \sum_{j\ge 1}\left ( 1 - \frac{1}{W^\dagger(x)} \right)^{j-1}\frac{1}{k^{j-1}}\sum_{i\ge j}  \left (1 -  \frac{1}{F^\dagger(mT)}\right)^{i-1} 
  G_{i-j}(x) \\
  =&\frac{\theta s_{m}}{F^\dagger(mT)}\int_0^{mT} dx \sum_{j\ge 1}\left ( 1 - \frac{1}{W^\dagger(x)} \right)^{j-1}\frac{1}{k^{j-1}}\left (1 -  \frac{1}{F^\dagger(mT)}\right)^{j-1} \sum_{i\ge 0}  \left (1 -  \frac{1}{F^\dagger(mT)}\right)^{i} 
  G_{i}(x) \\
  =&\frac{\theta s_{m}}{F^\dagger(mT)}\int_0^{mT} dx \sum_{j\ge 1}\left ( 1 - \frac{1}{F^\dagger(x)} \right)^{j-1}\frac{1}{k^{j-1}} \sum_{i\ge 0}  \left (1 -  \frac{1}{F^\dagger(mT)}\right)^{i} 
  G_{i}(x) \\
   =&\frac{\theta s_{m}}{F^\dagger(mT)}\int_0^{mT} dx  \frac{1}{1-\frac{1}k +\frac{1}{k F^\dagger (x) }} \sum_{i\ge 0}  \left (1 -  \frac{1}{F^\dagger(mT)}\right)^{i} 
  G_{i}(x) \\
\end{align*}

Now we write for any $i\ge 1$
$$
 \left (1 -  \frac{1}{F^\dagger(mT)}\right)^{i} 
  G_{i}(x)=: A^i (B(i+1)+C)
$$
where $A=1 -  1/F^\dagger(mT)$, $B=1/W^\dagger(x)^2$ and $C = 2/W^\dagger(x)-2/W^\dagger(x)^2$, so that
\begin{align*}
     \sum_{i\ge 0}  \left (1 -  \frac{1}{F^\dagger(mT)}\right)^{i} 
  G_{i}(x) =& 1+\sum_{i\ge 1}A^i (B(i+1)+C) \\
    =& 1-B+B\sum_{i\ge 0}(i+1)A^i+CA\sum_{i\ge 0}A^i\\
    =&1-B +\frac{B}{(1-A)^2}+\frac{CA}{1-A}\\
    =&1-B +BF^\dagger(mT)^2 +C(F^\dagger(mT)-1) \\
    =&1+\frac{F^\dagger(mT)^2-1}{W^\dagger(x)^2}+\frac{2(F^\dagger(mT)-1)}{W^\dagger(x)}\left(1-\frac1{W^\dagger(x)}\right)\\
   % =&\left(1-\frac{1}{W^\dagger(x)}+\frac{F^\dagger(mT)}{W^\dagger(x)}\right)^2 +1 - \left(1-\frac{1}{W^\dagger(x)}\right)^2 -\frac{1}{W^\dagger(x)^2}-\frac{2}{W^\dagger(x)}\left(1-\frac{1}{W^\dagger(x)}\right)\\
     =&\left(1-\frac{1}{W^\dagger(x)}+\frac{F^\dagger(mT)}{W^\dagger(x)}\right)^2\\
     =& \left(\frac{F^\dagger(mT)}{F^\dagger(x)}\right)^2.
\end{align*}
Hence we eventually get
$$
\mathcal S_{k,m,n} =  
 \theta  s_{m}F^\dagger(mT)\int_0^{mT}   \frac{k\ dx}{F^\dagger (x) (1+(k-1)F^\dagger (x))}
$$

\textbf{Conclusion}

Collecting all the terms we get: 
\begin{align*}
    M'_{k,m,n} &= \mathcal S_{k,m,n}  +  \mathcal C_{k,m,n} \\
   &= \theta s_m F^\dagger(mT) \left [ \int_0^{mT} \frac{k dx}{F^\dagger(x)(1+(k-1)F^\dagger(x))} +  \int_0^{nT} \frac{F(nT)}{F(x)}dx  \right ] \\
\end{align*}

\end{proof}
%---------------------------------------------------------------------------------------------

\section*{Artificial Selection of Droplets}

\begin{proof}[Proposition \ref{lem:timeToK} - Cumulative distribution functions]
\label{proof:timeToK}

\textbf{Time to accumulate $\Theta$ mutations}

Let \(\left ( e^{(i,j)}_{p_{ij}} \right )_{i,j \in \mathbb N^2
 }\) be independent exponential random variables with parameter
\(p_{ij}\). Without droplet selection, \(\Gamma\) is the minimum of a
set of sums of \iid exponential variables:
\begin{equation*}
\Gamma = \inf_{j = 1 \ldots D} \left ( \sum_{i=1}^\Theta  e^{(i,j)}_{\frac{\rho}{D}}  \right)_{j}
\end{equation*}

\textbf{With droplet selection}, \(\Gamma^*\) is a sum of a minimum of \iid
exponential variables, (i.e. a sum of exponential variables):
\begin{equation*}
\Gamma^* =  \sum_{i=1}^\Theta \left ( \inf_{j = 1 \ldots D}  e^{(i,j)}_{\frac{\rho}{D}} \right)_{i} =  \sum_{i=1}^\Theta  e^{(i)}_{\sum^D_{j=1} \frac{\rho}{D}} = \sum_{i=1}^\Theta  e^{(i)}_\rho,
\end{equation*}
the cumulative distribution function of \(\Gamma^*\) is the Erlang distribution with parameters $\Theta$ and $\rho$:
\begin{align*}
\mathbb P (\Gamma^* \leq x ) = 1 - e^{-\rho x} \left (\sum^{\Theta-1}_{u=0} \frac{(\rho x)^u}{u!} \right )\\
\end{align*}

Note that it does not depend on \(D\) any more: if the number of
droplets is in the order of one over the scaled invasion rate, the
time to accumulate \(\Theta\) mutations with droplet selection does not
depend on the invasion rate any more.

\textbf{Without droplet selection}, the cumulative distribution function of \(\Gamma\) is:

\begin{align*}
\mathbb P (\Gamma \leq x ) &= 1 - \mathbb P (\Gamma > x )\\
&= 1 - \mathbb P \left ( \bigcap_{j = 1}^D \left [ \left ( \sum_{i=1}^\Theta  e^{(i,j)}_{\frac{\rho}{D}}  \right)_{j}  > x \right ] \right)\\
&= 1 - \left [ \mathbb P \left ( \left ( \sum_{i=1}^\Theta  e^{(i,j)}_{\frac{\rho}{D}}  \right)_{j}  > x \right ) \right ]^D & \text{(The lineages are independent)} \\
&= 1 - \left [ 1 - \mathbb P \left ( \left ( \sum_{i=1}^\Theta  e^{(i,j)}_{\frac{\rho}{D}}  \right)_{j} \leq x \right ) \right ]^D \\
&= 1 - \left [ 1 - 1 + e^{-\frac{\rho}{D}x} \left (\sum^{\Theta-1}_{u=0} \frac{(\rho x)^u}{D^u u!} \right ) \right ]^D & \text{(sum of indep. exp. r.v.)}\\
&= 1 - \left [ e^{-\frac{\rho}{D}x} \left (\sum^{\Theta-1}_{u=0} \frac{(x \rho)^u}{D^u u!} \right ) \right ]^D\\
&= 1 - e^{-\rho x} \left ( \sum^{\Theta-1}_{u=0} \frac{(\rho x)^u}{D^u u!} \right )^D \\
\end{align*}

\bigskip
\textbf{Comparison of the two regimes}

Let us show that $\Gamma$ stochastically dominates $\Gamma^*$. Denote by $N^*$ the process counting the number of mutations in the selection regime. In the parallel regime, denote by $N^{(j)}$ the process counting the number of mutations in the $j$-th lineage. We know that $N^*$ is a Poisson process with intensity $\rho$, whereas the $N^{(j)}$ are independent Poisson processes with common intensity $\rho/D$, so that for all $t$
\begin{equation}
    N^*_t \stackrel{(d)}{=}\sum_{j=1}^D N^{(j)}_t \geq \max_{j=1\ldots D}N^{(j)}_t 
\end{equation}
Now note that $\Gamma^*$ is the first time when $N^*$ reaches $\Theta$, whereas $\Gamma$ is the first time when $\max_{j=1\ldots D}N^{(j)}$ reaches $\Theta$. Since the latter counting process is stochastically smaller than the former, we get the desired stochastic inequality.

Now by the law of large numbers, we have the a.s. convergence of $N^*_t/t$ to $\rho$ and for each $j$, of $N^{(j)}_t/t$ to $\rho/D$, which also yields the a.s. convergence of $\max_{j=1\ldots D}N^{(j)}_t/t$ to $\rho/D$. As a consequence, as $\Theta\to\infty$, $\Gamma^*/\Theta$ converges a.s. to $1/\rho$ and $\Gamma/\Theta$ converges a.s. to $D/\rho$. This proves that $\Gamma/\Gamma^*$ converges a.s. to $1/D$.  

\end{proof}

\end{document}